\documentclass[aps,pra,twocolumn,superscriptaddress,10pt]{revtex4-2}
\usepackage{graphicx}
\usepackage{dcolumn}
\usepackage{bm}
\usepackage{amsmath}
\usepackage{braket}
\usepackage{comment}
\usepackage{enumitem}
\usepackage{siunitx}
\usepackage{xcolor}
\usepackage[caption=false]{subfig}
\usepackage[colorlinks=true,linkcolor = blue,urlcolor=blue,citecolor = blue,anchorcolor = blue]{hyperref}
\usepackage{cleveref}

\begin{document}
\frenchspacing

\title{Self-Sustained Oscillations of a Nonlinear Optomechanical System in the Low-Excitation Regime}

\author{Shivangi Dhiman}
\thanks{These authors contributed equally to this work.}
\affiliation{Institute for Quantum Materials and Technology, Karlsruhe Institute of Technology, 76131 Karlsruhe, Germany}
\author{K. Rubenbauer}
\thanks{These authors contributed equally to this work.}
\affiliation{Walther-Meißner-Institut, Bayerische Akademie der Wissenschaften, 85748 Garching, Germany}
\affiliation{Physics Department, TUM School of Natural Sciences, Technical University of Munich, 85748 Garching, Germany}
\author{T. Luschmann}
\affiliation{Walther-Meißner-Institut, Bayerische Akademie der Wissenschaften, 85748 Garching, Germany}
\affiliation{Physics Department, TUM School of Natural Sciences, Technical University of Munich, 85748 Garching, Germany}
\affiliation{Munich Center for Quantum Science and Technology (MCQST), 80799 Munich, Germany}
\author{A. Marx}
\affiliation{Walther-Meißner-Institut, Bayerische Akademie der Wissenschaften, 85748 Garching, Germany}
\author{A. Metelmann}
\email{anja.metelmann@kit.edu}
\affiliation{Institute for Quantum Materials and Technology, Karlsruhe Institute of Technology, 76131 Karlsruhe, Germany}
\affiliation{Institute for Theory of Condensed Matter, Karlsruhe Institute of Technology, 76131 Karlsruhe, Germany}
\affiliation{Institut de Science et d’Ingénierie Supramoléculaires (ISIS, UMR7006), University of Strasbourg and CNRS, 67000 Strasbourg, France}
\author{H. Huebl}
\email{huebl@wmi.badw.de}
\affiliation{Walther-Meißner-Institut, Bayerische Akademie der Wissenschaften, 85748 Garching, Germany}
\affiliation{Physics Department, TUM School of Natural Sciences, Technical University of Munich, 85748 Garching, Germany}
\affiliation{Munich Center for Quantum Science and Technology (MCQST), 80799 Munich, Germany}

\date{\today}

\begin{abstract}
Manifesting across all time, mass and length scales, nonlinearities lie at the core of numerous physical phenomena. Next-generation quantum applications, such as quantum sensing, require the combination of nonlinearity with non-classical correlations. This necessitates the search for an experimental platform which enables a nonlinear response at ultra-low excitation levels in a system with practical sensing potential and quantum compatibility. Here, we report the observation and theoretical modeling of nonlinear dynamics in a mechanical system driven at the single-excitation level. We achieve this using a cavity-optomechanical platform with large single-photon coupling rates and a nonlinear microwave resonator. Specifically, the large Kerr nonlinearity of our superconducting microwave circuit reduces the threshold for the observation of nonlinear dynamics by four orders of magnitude, making this regime experimentally accessible at the few-photon level. The parameter-based quantitative predicative power of the theoretical description underlines our deep understanding of the physics involved and that this device concept paves the way for experiments with non-classical microwave drive schemes.
\end{abstract}

\maketitle

Nonlinear dynamics is fundamental to both classical and quantum systems. It gives rise to an extremely rich class of physics, like bifurcation \cite{hopf-bifurcation}, chaos \cite{bakemeier2015,lu2015}, and synchronization \cite{jensen1998,amitai2017}. Typically, these phenomena are highly sensitive to initial conditions, requiring a thorough study and deep understanding to make precise predictions and allow for an effective utilization of nonlinear dynamics. While nonlinear systems are ubiquitous in nature, their systematic experimental study often relies on classical mechanical systems or electronic circuits \cite{dykman-nonlinear-oscillators-2012-book,Strogatz-nonlin-dyn-chaos}. Nonlinearities also play a substantial role in quantum mechanics and are unmissable for quantum state generation \cite{yurke-cat-state-1986,Hofheinz-qstates-nat2009,Vlastakis-100photoncat-sci2013,puri-quantum-state-kerr-two-photon,he-fast-cat-kerr-tuning}.

Optomechanical systems parametrically couple an electromagnetic resonator with a mechanical mode, enabling extremely sensitive readout schemes for mechanical sensing applications \cite{aspelmeyer-cavity-optomechanics,Barzanjeh-opto-quantum-tech,Metcalfe-apl-cav-optomech-aprev2014,li-cav-optom-sens-nanophoto2021}. These systems allow for the monitoring of minute forces originating from e.g.~electromagnetic fields, Bose-Einstein condensates, mass, and even gravitational waves \cite{Brennecke-bose-einstein-science2008,ligo-virgo-gravitation-waves}. Their extreme sensitivity is linked to the ability to prepare the mechanical mode in the quantum mechanical ground state \cite{chan-ground-state-laser-cooling,OConnell-mech-ground-nat2010,Teufel-mech-ground-nat2011} or non-classical states like squeezed states \cite{Wollmann-mech-squeez-sci2015,Pirkkalainen-mech-squeez-prl2015,Barzanjeh-entangle-mech-nature2019,Lecocq-nondem-massive-prx2015,Barzanjeh-on-chip-circulator-natcom2017}. In addition, the same devices are also ideally suited for exploring nonlinear physics both in the classical and quantum domains \cite{Huber-squeez-mech-mode-prx2020,Ochs-freq-comb-mech-mode-prx2022,Kuang-phonon-laser-optomech-natphys2023}. However, due to the weak optomechanical interaction strength, these systems have been mostly limited to the classical regime up to this point.There have been numerous studies on such nonlinear dynamics in the classical domain, specifically self-sustained oscillations both from a theoretical and experimental perspective \cite{carmon-selfosc-2007,marquardt2006,krause2015,srdas2023,roque2020}. Most of the works rely on driving the system at high input powers to observe these features. In general, a combination of non-classicality and nonlinear response of the mechanical system can be accessed in the single-photon strong-coupling regime, where the optomechanical coupling rate exceeds the photonic and phononic decay rate. In addition to accessing nonlinear quantum dynamics, this regime allows for the generation of arbitrary quantum states \cite{Clerk-phonon-shot-noise-prl2010,Liu-phonon-block-pra2010,Nunnenkamp-single-photon-prl2011,Rabl-photon-block-prl2011,Qian-strong-coupl-prl2012,Kronwald-omit-nonlinear-prl2013,Pdnation2013,Hauer-QND-single-photon-pra2018,Hauer-cool-to-cat-2023}.

Since entering the single-photon strong-coupling regime is experimentally demanding, suitable schemes need to be explored allowing to reduce the demands on the coupling strength. Promising strategies include enhancing the nonlinearity induced by the optomechanical coupling strength, e.g.~by implementing nonlinear elements in the electromagnetic circuit. Inductively coupled cavity-optomechanical systems using superconducting microwave (quantum) circuits with embedded mechanical elements, as demonstrated in Refs.~\cite{schmidt-first-paper,rodrigues-first-nanoelectromech-device,zoepfl-nonlinear-cooling,Bera-singh-device-comphys2021}, are ideally suited to test this conjecture. These systems combine an intrinsic nonlinear microwave resonator with a mechanical element and the ability to prepare the mechanical modes close to the ground state \cite{zoepfl-nonlinear-cooling,nico-nonlinear-cooling,Bothner-4wave-cool-comphys2022,Majumder-cooling-qbit-mech-prr2022,deeg-lukas-pra012025-backaction-bistable}.

Here, we experimentally explore and quantitatively model an optomechanical system with a strong nonlinearity implemented as a superconducting quantum circuit. We observe self-sustained oscillations \cite{carmon-selfosc-2007,marquardt2006,krause2015,srdas2023} of the mechanical oscillator by measuring the scattering response of the nonlinear cavity at excitation levels as low as a few photons. Our semi-classical model quantitatively describes the observed nonlinear response using independently determined system parameters. Notably, the excitation level of the microwave resonator required to trigger self-sustained mechanical oscillations is by several orders of magnitude lower than in conventional optomechanical systems based on linear cavities. The observation of nonlinear features at occupations as low as few photons makes the setup compatible with future studies in the quantum domain. Specifically, this provides a pathway to investigate nonlinear dynamics driven by non-classical excitation schemes.

\begin{figure}[t]
    \includegraphics{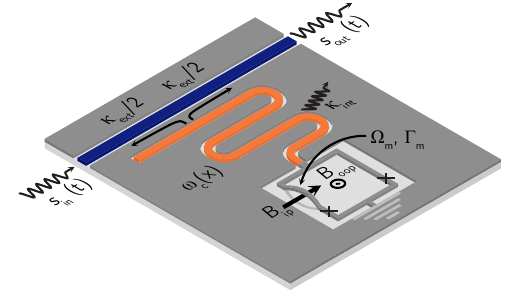}
    \caption{Schematic representation of the device featuring a superconducting flux-tunable $\lambda/4$ coplanar waveguide (CPW) resonator (orange), which is shunted to ground via a dc-SQUID. Signal input $s_\mathrm{in} \left( t \right)$ and output $s_\mathrm{out} \left( t \right)$ are achieved through a transmission line (blue). The dc-SQUID is partially suspended, forming two nanomechanical string resonators (only one of which is depicted here). The frequency of the microwave resonator is tunable via an out-of-plane magnetic field $B_\mathrm{oop}$. The application of an in-plane magnetic field $B_\mathrm{ip}$ induces an optomechanical coupling between the out-of-plane displacement of the nanostring and the CPW resonator. The sample is probed at a temperature of $\SI{70}{\milli\kelvin}$ in a commercial dilution refrigerator. Details on the device fabrication and cryogenic wiring can be found in Sections \ref{SI:sec:fab-layout} and \ref{SI:sec:mw-setup} of the SI, respectively.}
    \label{main:fig:1-setup}
\end{figure}

\textbf{Device concept.} We use the inductively coupled nano-electromechanical device, which has been described in Refs.~\cite{schmidt-first-paper,luschmann-mech-freq-control}. The microwave cavity is frequency tunable between $6.6$ and $\SI{7.35}{GHz}$ by an external magnetic field. It is realized as a superconducting $\lambda/4$ coplanar waveguide (CPW) resonator made from an aluminum thin film (see Fig.~\ref{main:fig:1-setup}), which is shunted to ground via a dc superconducting quantum interference device (dc-SQUID). The dc-SQUID is partially suspended and hosts two nanomechanical string resonators with a length of $\SI{30}{\micro\meter}$ and resonance frequencies of $\Omega_\mathrm{m} / \left(2\pi\right) \approx \SI{5.6}{\mega\hertz}$, sharing loss rates of approximately $\Gamma_\mathrm{m} / \left(2\pi\right) \approx \SI{10}{\hertz}$. It adds a magnetic flux-dependent inductance to the circuit, which allows for tuning of the microwave resonance frequency $\omega_\mathrm{c}$, controls the optomechanical single-photon coupling rate $g_0$, and adds a Kerr nonlinearity $\mathcal{K}$ to the microwave harmonic oscillator. We adjust the applied flux $\Phi$ and the total magnetic field $B_\mathrm{ext}$ using an in-plane and an out-of-plane magnetic field $B_\mathrm{ip}$ and $B_\mathrm{oop}$, respectively. In this paper, we focus on the mechanical mode at $\Omega_\mathrm{m}/(2\pi)=\SI{5.607716}{MHz}$, with large single-photon optomechanical coupling rates $g_0$ in the range of a few to tens of kilohertz.

This electromechanical device can be accurately modeled using the Hamiltonian for a driven optomechanical system with the addition of a Kerr nonlinearity to the cavity mode \cite{nation2008,zoepfl-nonlinear-cooling,nico-nonlinear-cooling},
\begin{equation} \label{main:eq:hamilton}
    \hat{H} / \hbar = -\Delta \hat{a}^{\dag} \hat{a} + \Omega_\mathrm{m} \hat{b}^{\dag} \hat{b} -\frac{\mathcal{K}}{2} \hat{a}^{\dag} \hat{a}^{\dag} \hat{a} \hat{a} + g_0 \hat{a}^{\dag} \hat{a}(\hat{b} + \hat{b}^{\dag}).
\end{equation}
Here, $\hat{a}$($\hat{b}$), $\hat{a}^{\dag}$($\hat{b}^{\dag}$) are the annihilation and creation operators for the cavity (mechanical) mode, and $\Delta = \omega_\mathrm{p} - \omega_\mathrm{c}$ is the detuning in the frame rotating with respect to the external microwave probe frequency $\omega_\mathrm{p}$. Note that $\omega_c$ is the resonance frequency of the nonlinear cavity at very small driving strength, i.e.~in its linear regime.

\textbf{Stability analysis.} Since self-sustained oscillations are a signature of instability in the mechanical mode, we first discuss the stability of our system as a function of the detuning $\Delta$ and the input photon flux $n_\mathrm{in}$. For this analysis, we solve for the fixed points of the system and perform a linear stability test in their proximity \cite{Strogatz-nonlin-dyn-chaos} (see Section \ref{si:phasediagram} of the SI for more details). Generally, such an optomechanical system has one or three real solutions for fixed points \cite{roque2020} depending on $n_{\mathrm{in}}$ and $\Delta$. This is because the number of physical (real) solutions for the fixed points is governed by a cubic equation, which can either have one or three real solutions depending on the parameters. If the real part of all the eigenvalues of the stability (drift) matrix is negative, then the corresponding fixed point is stable.

\begin{figure}[t]
    \centering
    \includegraphics{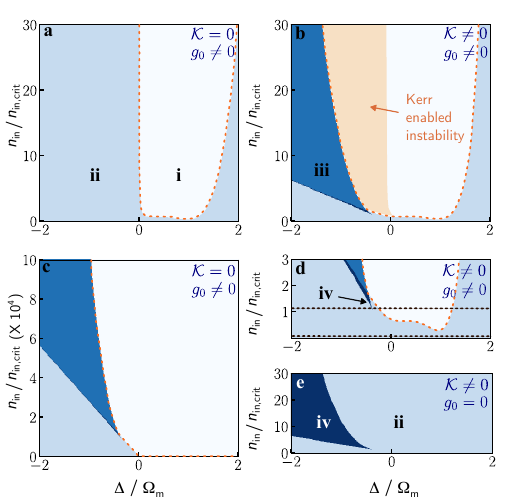}
    \caption{Stability diagram for an optomechanical system as described by Eq.~(\ref{main:eq:hamilton}) as a function of the normalized detuning of the probe frequency from the cavity resonance, $\Delta / \Omega_\mathrm{m}$, and the input photon flux $n_\mathrm{in}$ normalized with respect to the critical photon flux $n_{\mathrm{in,crit}}$ at which the optomechanical system becomes bistable. On the left, panels \textbf{a} and \textbf{c} show a linear system with $\mathcal{K}=0$, whereas in panels \textbf{b}, \textbf{d}, and \textbf{e} on the right, a Kerr nonlinearity of $\mathcal{K} / 2 \pi = \SI{70}{\kilo\hertz}$ is assumed. For all panels with non-zero coupling, $g_0 = \SI{4.69}{\kilo\hertz}$. $n_{\mathrm{in,crit}}$ is the same in all panels and calculated for the case of $\mathcal{K} = \SI{70}{\kilo\hertz}$. It corresponds to an input power of $\SI{-120}{dBm}$, where the cavity starts to bifurcate at an occupation of $\bar{n}_\mathrm{c,crit} = 19$. These parameter values correspond to parameter set III in the experiment (see table \ref{tab:param-sets-main}). The colors depict different stability regions with different amounts of (un)stable solutions. In region \textbf{i}, we find one unstable, in \textbf{ii} one stable, in \textbf{iii} one stable and two unstable, and in \textbf{iv} two stable and one unstable fixed point. Unstable regions are encircled by a dashed orange line. In \textbf{b}, the nonlinearity of the cavity causes a Kerr-enhanced instability, which extends the unstable region compared to a linear system (orange shaded area). Note that panel \textbf{c} extends \textbf{a} to vastly larger input powers. Only then one can observe multi-stability in a linear system, whereas our Kerr nonlinear cavity reduces the input power required to reach this regime by four orders of magnitude. Panel \textbf{d} is a zoom into the low power region of \textbf{b}, where the dashed horizontal lines represent the input powers displayed in Fig.~\ref{main:fig3:s21comp}c,d. Panel \textbf{e} shows the phase diagram for a classical nonlinear duffing resonator with zero optomechanical interaction $g_0 = 0$.}
    \label{main:fig2:stability-dia}
\end{figure}

Figure \ref{main:fig2:stability-dia} compares the resulting stability diagram of our electromechanical system with and without a non-zero Kerr and optomechanical interaction, where we normalized the input photon flux in all cases to the same critical photon flux $n_{\mathrm{in,crit}} = 2 \kappa^3 / (3 \sqrt{3} \kappa_{\text{ext}} \mathcal{K}_{\text{eff}})$ calculated for $\mathcal{K}, g_\mathrm{0} \neq 0$. Here, the effective Kerr constant $\mathcal{K}_{\mathrm{eff}} = \mathcal{K} + \mathcal{K}_{\mathrm{m}}$ is composed of the intrinsic nonlinearity of the superconducting circuit $\mathcal{K}$ and the nonlinearity originating from the radiation pressure of the mechanical mode $\mathcal{K}_{\mathrm{m}}$ \cite{aldana2013,nico-nonlinear-cooling}. While the case of of a linear cavity, $\mathcal{K}=0$ in Fig.~\ref{main:fig2:stability-dia}a,c, shares features like multi-stability with the one of a nonlinear cavity, $\mathcal{K}\neq0$ in Fig.~\ref{main:fig2:stability-dia}b,d,e, a non-zero $\mathcal{K}$ dramatically reduces the excitation powers required to enter this regime. In particular, the non-zero Kerr nonlinearity shifts the transition from at least one stable (regions \textbf{ii-iv}) to one unstable (region \textbf{i}) fixed point towards negative detunings (i.e. the red sideband) for very low intra-cavity photon occupations. Because the cavity resonance shifts towards negative detunings due to the strong Kerr nonlinearity, the effective sidebands also shift with power, enabling an effective heating of the mechanical mode at negative detunings. Moreover, the multi-stable area (regions \textbf{iii,iv}), with three fixed points is accessible at several orders of magnitude lower input powers when compared to a linear optomechanical system. We can understand this by the critical photon flux, which quantifies the boundary where our system becomes multistable, which is a nonlinear phenomenon. The critical photon flux $n_{\mathrm{in,crit}}$ is determined by the effective Kerr $\mathcal{K}_{\mathrm{eff}}$. For our system parameters in Fig.~\ref{main:fig2:stability-dia}, $\mathcal{K}_\mathrm{m} \approx 10^{-4} \mathcal{K}$, which means that the critical input power is roughly four orders of magnitude smaller compared to a linear cavity. Hence, the finite Kerr nonlinearity qualitatively alters the stability diagram as displayed in Fig.~\ref{main:fig2:stability-dia} a,b. For large positive detunings, the system is forced back into stability for $\mathcal{K} \neq 0$ by the Kerr nonlinearity, whereas for $\mathcal{K} = 0$ this boundary is only determined by $\Delta$ and $g_0$. By setting $g_0 = 0$ as in Figure \ref{main:fig2:stability-dia}e, we recover the phase diagram of a classical Kerr nonlinear resonator with three fixed points (two stable and one unstable, region \textbf{iv}).

We can intuitively understand the qualitative form of the stability diagram. The region with one unstable fixed point links to the possibility of stimulating self-sustained oscillations of the mechanical system using a blue sideband drive (region \textbf{i}). The multiple fixed points originate from the total Kerr nonlinearity of the system with contributions from the microwave circuit and the optomechanical interaction \cite{nico-nonlinear-cooling}.

Within the stability diagrams shown in Fig.~\ref{main:fig2:stability-dia} we can identify three types of bifurcations: i) a Hopf bifurcation at the transition between a stable and an unstable fixed point (boundary between region \textbf{i} and \textbf{ii}) and the same transition when involving a total of three fixed points (transition from region \textbf{iv} to \textbf{iii}); ii) a (inverse) saddle node bifurcation occurring between areas with three  and a single fixed point (transition from \textbf{ii} to \textbf{iv}, and from \textbf{iii} to \textbf{i}). More details can be found in Section \ref{si:phasediagram} of the SI.

\textbf{Cavity scattering response.} To link the stability diagram to the experimental data, we model the scattering response of the microwave resonator $S_{21} = s_\mathrm{out}/s_\mathrm{in}$ in different stability regimes using two approaches: an approximative analytical model and the full numerical solution of the equations of motion of the system, both of which are discussed in Section \ref{si:sec:theory-model} of the SI. Both approaches employ standard input-output theory \cite{gardiner1985} to obtain the cavity response. For the analytical description, we distinguish between the response obtained in the stable and unstable regions. When the system is stable, the scattering response $S_{21}$ is given by
\begin{equation}
    S_{21} = 1 -\frac{\kappa_{\mathrm{ext}}}{2} \frac{1}{-i(\Delta+\mathcal{K}_{\mathrm{eff}} \bar{n}_{\mathrm{c}}) + \kappa/2} .
    \label{main:eq:S21-stable-region}
\end{equation}
Here, $\kappa = \kappa_{\text{ext}}+\kappa_{\text{int}}$ is the total loss rate of the microwave resonator with the coupling rate $\kappa_{\text{ext}}$ and the internal loss rate $\kappa_{\text{int}}$, and $\bar{n}_\mathrm{c}$ is the steady state photon occupation number of the microwave cavity. 

In the regime of instability, we solve the classical equations of motion assuming a coherent oscillation of the mechanical mode with $\beta = \bar{\beta}+B'e^{-i\phi}e^{-i\Omega_{\mathrm{m}} t}$, where $\bar{\beta}$ represents the static displacement of the mechanical resonator and $B'$ ($\phi$) are the amplitude (phase) associated with the oscillations of the mechanical displacement \cite{marquardt2006,krause2015}. $B'$ is determined via the power balance condition $\Gamma_\mathrm{m} + \Gamma_{\mathrm{opt}} = 0$ with the optomechanical damping rate $\Gamma_{\mathrm{opt}}$\cite{Ludwig2008}. The scattering response in this regime is 
\begin{equation}
    S_{21} = 1 - \frac{\kappa_{\mathrm{ext}}}{2} \sum_{n} \frac{J_n(z_1) J_n(z_1)}{-i(\Delta+\mathcal{K}_{\mathrm{eff}} \bar{n}_\mathrm{c} + n\Omega_\mathrm{m}) + \kappa/2} ,
    \label{main:eq:S21-instable-region}
\end{equation}
with $J_n$ the Bessel function of the first kind and $z_1 = 2B' g_0/\Omega_{\mathrm{m}}$.

In addition, we simulate the response of the system by numerically solving the equations of motion for our Hamiltonian and employing the input-output relation to obtain the scattering response. The simulation fully captures the dynamic evolution of the correct photon number without assuming a steady state, as done in the analytical model. This allows for the correct incorporation of the effect from the Kerr nonlinearity, especially at the transition boundary from stable to unstable dynamics. Furthermore, we can precisely predict the time required for the system to reach its steady state (see section \ref{si:simulation} of the SI for more details). Note that we have restricted the discussion to classical dynamics and thus have not investigated quantum fluctuations since our single tone measurement only gives access to the mean field scattering response of the system.

\textbf{Pulsed measurement technique to avoid transient dynamics.} For the experiment, we apply a combination of in-plane and out-of-plane magnetic fields. This sets the single-photon coupling rate\linebreak $g_0 = \left( \partial_\Phi\omega_\mathrm{c} \right) \gamma B_\mathrm{ext} x_\mathrm{zpf} l$\cite{schmidt-first-paper, rodrigues-first-nanoelectromech-device} and the intrinsic Kerr nonlinearity $\mathcal{K}$ of the cavity (see table \ref{tab:param-sets-main}, and SI). Here, $\partial_\Phi\omega_\mathrm{c}=\partial \omega_\mathrm{c} / \partial \Phi$ is the flux responsivity, i.e. the derivative of the cavity resonance frequency with respect to the applied flux $\Phi$, $\gamma$ is the mode shape factor, and $l$ is the length of the nanostring.

Specifically, we set $B_\mathrm{ip} = \SI{30}{\milli\tesla}$ for all experiments presented and control $g_0$ between zero and several tens of kilohertz using $\Phi_\mathrm{oop}$ via $\partial_\Phi \omega_\mathrm{c} \approx \partial_{\Phi_\mathrm{oop}} \omega_\mathrm{c}$. Notably, $\Phi_\mathrm{oop}$ also affects the Kerr nonlinearity $\mathcal{K}$ of the cavity. Thus, by adjusting the out-of-plane magnetic field $B_\mathrm{oop}$ or $\Phi_\mathrm{oop}$, we are able to explore a wide range of system parameters (see table \ref{tab:param-sets-main}). To demonstrate our quantitative understanding of the response, we determine the system parameters for four distinct working points using independent techniques as discussed in Section \ref{SI:sec:device-params} of the SI. Those then act as input parameters for our theoretical model, which we then compare to the experimental data.

\begin{table}[b]
\caption{\label{tab:param-sets-main}
Experimental system parameters for the working points discussed in the main text. Additional parameter sets are discussed in Section \ref{SI:sec:add-param-set} of the SI. The parameters are determined independently and used as input for the analytical model and to numerically simulate the scattering response. Since the actual occupation of the nonlinear microwave cavity depends on the detuning $\Delta$, we give the cavity occupation at bifurcation and at the critical detuning $\Delta_\mathrm{crit}$ as $\bar{n}_\mathrm{c,crit} = \kappa / \sqrt{3} \mathcal{K}_\mathrm{eff}$ as a reference. Note that due to the tunability of our microwave cavity, we are able to cover substantially different values of $g_0$, $\mathcal{K}$, and thus also the ratio $g_0 / \mathcal{K}$.}
\begin{ruledtabular}
\begin{tabular}{llll}
 & Set II & Set III & Set IV \\
\hline 
$\omega_\mathrm{c} / 2 \pi$ (GHz) & $7.310$ & $7.241$ & $7.006$ \\
$\kappa_\mathrm{int} / 2 \pi$ (MHz) & $0.60$ & $0.68$ & $2.33$ \\
$\kappa_\mathrm{ext} / 2 \pi$ (MHz) & $1.72$ & $1.64$ & $1.55$ \\
$\mathcal{K} / 2 \pi$ (kHz) & $20$ & $70$ & $1.4 \times 10^3$ \\
$\bar{n}_\mathrm{c,crit}$ & $67$ & $19$ & $1.6$ \\
$\Omega_\mathrm{m} / 2 \pi$ (MHz) & $5.607 653$ & $5.607 483$ & $5.607 110$ \\
$\Gamma_\mathrm{m} / 2 \pi$ (Hz) & $14$ & $12$ & $6$ \\
$g_0 / 2 \pi$ (kHz) & $1.95$ & $4.69$ & $18.4$ \\
$g_0 / \mathcal{K}$ & $0.10$ & $0.07$ & $0.01$ \\
\end{tabular}
\end{ruledtabular}
\end{table}

\begin{figure}[t]
    \centering
    \includegraphics{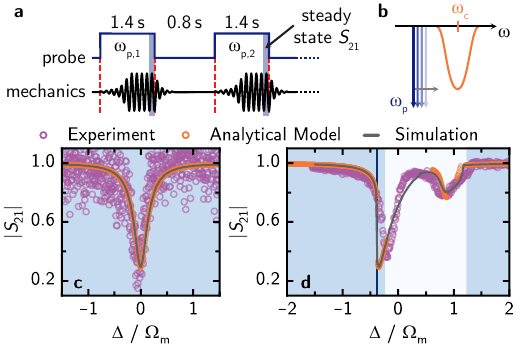}
    \caption{\textbf{a}, Measurement protocol in the time domain: A probe signal at a given frequency $\omega_{\mathrm{p},1}$ is applied sufficiently long for the mechanical oscillator to ring up and reach its steady state. A waiting period follows, which allows the excitation to ring down completely. We repeat this for different frequencies $\omega_{\mathrm{p},2}$, $\omega_{\mathrm{p},3}$, ...~of the probe signal. The detuning $\Delta$ is calculated for each probe frequency with respect to the resonance frequency $\omega_\mathrm{c}$ of the cavity in the linear regime, i.e.~$\Delta = \omega_{\mathrm{p},i} - \omega_\mathrm{c}$ for $i = 1,2,3,...$. \textbf{b}, Measurement protocol in the frequency domain: A single microwave probe/pump tone with frequency $\omega_\mathrm{p}$ is swept around the cavity resonance $\omega_\mathrm{c}$. We measure its complex scattering parameter $S_\mathrm{21} \propto s_\mathrm{out} / s_\mathrm{in}$. \textbf{c} and \textbf{d}, Scattering response $|S_{21}|$ observed in the experiment (purple dots), calculated with our analytical model (orange dots) and obtained from numerical simulation (gray lines) for parameter set III and an input power of $n_\mathrm{in} / n_{\mathrm{in,crit}} = 0.008$ and $1.10$ corresponding to $P_\mathrm{d} = \SI{-139.6}{dBm}$ and $\SI{-118.6}{dBm}$, respectively. The different stability regions are highlighted in the background by the color code introduced in Fig.~\ref{main:fig2:stability-dia}. We see that both the analytical model and the simulation capture the nonlinear features, including the self-sustained oscillations observed as an additional absorption dip around $\Delta / \Omega_\mathrm{m} = 1$ in \textbf{d}.}
    \label{main:fig3:s21comp}
\end{figure}

To explore the stability diagram and the dynamic response of the system experimentally, we apply a single fixed frequency microwave tone at $\omega_\mathrm{p}$ to the system ($s_\mathrm{in}(t)$) and record the response in the time domain ($s_\mathrm{out}(t)$) (see Figs.\ \ref{main:fig3:s21comp}a,b and \ref{main:fig:1-setup}) in the form of a time-dependent complex transmission parameter $S_\mathrm{21} \left( t \right)$. The tone is applied, and $S_\mathrm{21}(t)$ is recorded for $\SI{1.4}{\second}$ to experimentally measure the ring-up of the mechanics and verify that the steady state is reached. To display the transmission parameter only in the steady state, $S_\mathrm{21}(t)$ is averaged over the last $\SI{2}{\milli\second}$ of the pulse. This ensures that it is not altered by transient or ring-up effects, which occur initially after the probe pulse is applied. In between subsequent interrogations, we allow the system to equilibrate with its thermal environment for $\SI{0.8}{\second}$ so that any excitation of the mechanics fully depletes before the next pulse is applied. We repeat this procedure as a function of the frequency of the microwave tone $\omega_\mathrm{p}$. A full $S_\mathrm{21}(\omega,t)$ spectrum can be found section \ref{SI:sec:pulsed-transient} of the SI. Figure \ref{main:fig3:s21comp}c,d show resulting the steady state transmission as a function of frequency for two different excitation powers $P_\mathrm{d}$ at the cavity. This measurement protocol is essential to compare the experimental data with theory. Since our system is weakly dissipative ($\kappa < \Omega_\mathrm{m}$) along with $g_0 <\Omega_\mathrm{m}$, it can require up to $\SI{1}{\second}$ to reach steady state after the probe tone is applied, which we observe both in the simulation and the experiment. Hence, conventional frequency swept experiments can result in a transient behavior, which is beyond the scope of our theoretical treatment (for more details, see Sec.~\ref{SI:sec:pulsed-transient} of the SI). We choose this measurement protocol since it is a conceptually simple and fast measurement with good signal-noise at the given bandwidths and powers. Moreover, it allows for a direct comparison with our theoretical framework. A measurement of the self-sustained oscillations by the temporal behavior of the microwave output power, as done in, e.g., Ref.\,\cite{carmon-selfosc-2007}, is impractical due to the low power levels in our study.

\begin{figure*}[t]
    \centering
    \includegraphics{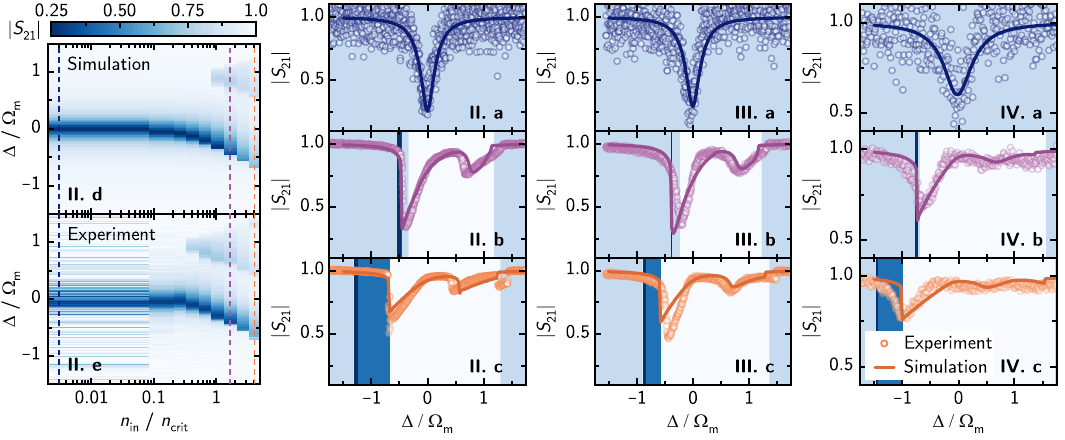}
    \caption{Power and parameter dependence of the cavity scattering response. The panels are labeled with capital Roman numbers according to the parameter sets, so that, e.g.~the leftmost two columns \textbf{II.a,b,c,d,e} refer to parameter set II in table \ref{tab:param-sets-main}. The two color plots \textbf{II.d,e} compare simulation and experiment for a wide set of input powers. The linecuts in \textbf{II.a,b,c} show the experimental data (points) and numerical simulation (lines) for a single input power increasing from top to bottom. The powers are indicated by the dashed vertical lines in \textbf{II.d,e}, and correspond to $n_\mathrm{in} / n_{\mathrm{in,crit}} = 0.003$, $1.63$, and $4.09$, translating to $P_\mathrm{d} = \SI{-139.6}{dBm}$, $\SI{-111.6}{dBm}$, and $\SI{-107.6}{dBm}$. For \textbf{III.a,b,c}, where both $g_0$ and $\mathcal{K}$ are increased, we show $n_\mathrm{in} / n_{\mathrm{in,crit}} = 0.008$, $1.10$, and $2.77$ ($P_\mathrm{d} = \SI{-139.6}{dBm}$, $\SI{-118.6}{dBm}$, and $\SI{-114.6}{dBm}$). Since \textbf{IV.a,b,c} on the right has the largest single-photon coupling $g_0$ and Kerr nonlinearity $\mathcal{K}$, the lowest absolute input powers are required for the additional absorption dips to appear on the blue sideband. Input powers of $n_\mathrm{in} / n_{\mathrm{in,crit}} = 0.04$, $1.45$, and $2.88$ are depicted, which correspond to $P_\mathrm{d} = \SI{-139.6}{dBm}$, $\SI{-123.6}{dBm}$, and $\SI{-120.6}{dBm}$. For all linecuts, the background is shaded according to the color code introduced with the stability diagram in Fig.~\ref{main:fig2:stability-dia}.}
    \label{main:fig4:colormap-slices}
\end{figure*}

\textbf{Connecting experiment, model and stability analysis.} Figure \ref{main:fig3:s21comp}c,d show the steady state transmission data for $P_\mathrm{d} = -139.6$ and $\SI{-118.6}{dBm}$, which correspond to $n_\mathrm{in} / n_{\mathrm{in,crit}} = 0.008$ and $1.10$, respectively. The power at the sample is calibrated as detailed in Section \ref{SI:sec:n-photon-calib} of the SI. The data recorded with the low excitation power shows the expected linear response of the microwave resonator. Surprisingly, even a moderate excitation level of only $P_\mathrm{d} = \SI{-118.6}{dBm}$ qualitatively alters the transmission signature in two ways by (i) deforming the main resonance and shifting it to negative detunings and (ii) the appearance of an additional absorption signature of the microwave cavity at frequencies corresponding to the blue sideband around $\Delta/\Omega_\mathrm{m} = 1$.

We quantitatively understand both excitation scenarios analytically using Eqs.~\ref{main:eq:S21-stable-region} and \ref{main:eq:S21-instable-region} to compute the scattering response. In the first case shown in Fig.~\ref{main:fig3:s21comp}c, the input is much below the critical input power $n_{\mathrm{in,crit}}$ such that the system is completely stable and can be correctly described using Eq.~\ref{main:eq:S21-stable-region}. This allows us to assert that we only have a single stable fixed point and thus only one photon branch exists for all detunings (corresponding to region \textbf{ii}).

For higher powers as in Fig.~\ref{main:fig3:s21comp}d, the analytically modeled response is piecewise combined using the stability diagram shown in Fig.~\ref{main:fig2:stability-dia}d. The scattering response in regions \textbf{ii-iv} with at least one stable fixed point is computed using Eq.~\ref{main:eq:S21-stable-region}, which already includes the shift of the transmission minimum due to the finite Kerr nonlinearity. Transitioning into region \textbf{i}, the system undergoes a Hopf bifurcation, which characterizes the onset of self-sustained oscillations of the mechanical mode\cite{Strogatz-nonlin-dyn-chaos}, hence we model the scattering response in this region with Eq.~\ref{main:eq:S21-instable-region}. In this range, we find further absorption signatures at multiples of the mechanical resonance frequency and frequencies higher than the response of the microwave cavity.

Notably, this spectral signature shows a Duffing-like shape similar to the main resonance, as it is slightly skewed and shifted from the first mechanical sideband to lower frequencies. This can be attributed to the Kerr nonlinearity again and is correctly reflected in our model. In addition, we operate in Fig.~\ref{main:fig3:s21comp}d above bifurcation, causing the photon branches to split into three. This corresponds to three fixed points, as in regions \textbf{iii,iv}, similar to the three photon branches of a Duffing oscillator. For smaller values of $g_0$, there are two stable (upper/lower) and one unstable (middle) photon branch as indicated by region \textbf{iv}. However, our system settles for nearly all detunings into the lower photon branch since the instability of the mechanical subsystem makes the upper branch inaccessible, corresponding to region \textbf{iii} with only one stable and two unstable fixed points. For more details, we refer to Section \ref{si:phasediagram} of the SI.

While our analytical model provides a good quantitative understanding of the system, it inherently includes approximations such as the assumption of a stationary state of the cavity, or neglects the excitation of the mechanical mode by higher-order sidebands. In addition, these constrain the model to finite frequency sections. Therefore, we also numerically solve the equations of motion to obtain the scattering response displayed in Fig.~\ref{main:fig3:s21comp}c,d, and find excellent quantitative agreement between experiment, analytical model, and simulation.

\textbf{Extending to various powers and system parameters.} Figure \ref{main:fig4:colormap-slices} compares the input power dependent spectral data with the numerical model for the parameter sets given in table \ref{tab:param-sets-main}, demonstrating our exquisite quantitative understanding. For these datasets, we repeat our transmission measurement for different input powers and detunings and display the steady state results as $|S_{21}|$. As predicted by the stability diagram and Eqs.~\ref{main:eq:S21-stable-region} and \ref{main:eq:S21-instable-region}, we observe a shift of the microwave resonator frequency to lower frequencies combined with a distorted Duffing-like lineshape for all three sets of parameters. In addition, we find a second absorption dip in the blue detuned regime, which shows a threshold behavior with the input power. As before, we attribute this feature to self-sustained oscillations of the mechanical mode observed through the cavity response and created by the effective blue sideband drive of the applied tone on the cavity. Thus, they reflect the instability of the system at these driving powers. This unique feature of our optomechanical system is linked to the comparably large $g_0$ and $\mathcal{K}$ values. All features of the spectra and their power dependence are in excellent agreement with the parameter-based and fit-free simulation. Line cuts at different input powers (see Fig.~\ref{main:fig4:colormap-slices}a,b,c for set II, III, and IV) underline this aspect. Only at very high input powers do we notice a slight deviation in absorption strength. This deviation might arise due to various reasons. First, the microwave drive induces a thermal occupation of the system that is not accounted for in our model. Second, the systems parameters might change at high input powers. In particular, a reduction of the loss rate $\kappa$ of the cavity due to the saturation of two-level systems can quickly influence the dip shape. Third, uncertainties in the power calibration might have a larger influence at larger drive powers. Lastly, higher-order nonlinearities, which are not included in our theoretical model, could also come into play at these powers.

Furthermore, the fixed point analysis quantitatively maps out the areas with different types and amounts of fixed points. The corresponding boundaries between stable and unstable regions are reflected experimentally, for example, as the sudden jump observed at the high frequency end of the sidedip around $\Delta / \Omega_\mathrm{m} \approx 1.4$ in Fig.~\ref{main:fig4:colormap-slices}II.c and III.c. Hence, for larger positive detunings, we do not observe and expect additional absorption features as our system remains stable.

Additional datasets presented in the SI underline the quantitative agreement between experiment and model. Between these, the changing single-photon optomechanical coupling $g_0$ shifts the instability barrier. For higher $g_0$ the sidebands appear already at lower input powers. This shows that, even though both the optomechanical coupling $g_0$ and the Kerr nonlinearity $\mathcal{K}$ strongly influence the stability regions and steady state times, we can model the nonlinear response to great quantitative agreement. Note that due to the large Kerr, the photon number in the cavity at which we observe these strongly nonlinear physics is in the range of only a few to tens of photons ($\bar{n}_\mathrm{c} \approx 2 \text{-} 120$).

\textbf{Discussion and Outlook.} In conclusion, we have observed self-sustained oscillations in a nonlinear nano-electromechanical device and demonstrated excellent quantitative understanding between our theory and experiment. While our theoretical analysis is restricted to the classical domain, the presence of such nonlinear features like limit cycles or period doubling \cite{srdas2023} marks the threshold for the parameter space where nonlinear interactions in the system become dominant. There are theoretical proposals predicting non-classical states in such regimes \cite{Pdnation2013}, but they require the still experimentally challenging single-photon strong coupling. Notably, the introduction of a nonlinear microwave resonator into our nano-electromechanical system drastically reduces the threshold for nonlinear dynamics to the few-photon level and hereby enables a potential pathway for the generation of non-classical states and the study of their dynamics in the nonlinear regime (see Section \ref{SI:sec:compare-other-works} of the SI). Both require the mechanical mode to be close to or in its ground state, which can be achieved by red sideband cooling. Additionally, the Kerr nonlinearity can be engineered as desired to explore potential pathways for state generation. Beyond the fundamental aspects, potential novel applications in quantum sensing envisage the nonlinear dynamics of non-classical states as a resource \cite{reviewdegen2017}, which come in reach given our findings combined with the versatile toolbox of superconducting quantum circuits.

\section*{Data availability}

The data displayed that support the findings of this study is openly available at the following url/doi: \url{10.5281/zenodo.18954807}.

\section*{Code availability}
The code used for numerical simulation is openly available at the following url/doi: \url{10.5281/zenodo.18954807}. The code is also made available on github \url{https://github.com/dhiman-shivangi/self-oscillations-optomechanics-codes}.

\section*{Acknowledgements}
We acknowledge funding from the Horizon Europe 2021–2027 Framework Programme under the Grant Agreement No. 101080143 (SuperMeQ) and from the Deutsche Forschungsgemeinschaft (DFG, German Research Foundation) under Germany's Excellence Strategy—EXC-2111-390814868. Further, this research is part of the Munich Quantum Valley, which is supported by the Bavarian state government with funds from the Hightech Agenda Bayern Plus.

\section*{Author contributions}
SD with AMe developed the theoretical model including its numerical solution. KR and TL performed the experiment and the experimental data analysis. HH, TL, and AMa devised the experiment. SD, KR, AMe, and HH wrote the manuscript with help from all authors.

\section*{Competing interests}
The authors declare no competing interests.

\bibliography{bibliography.bib}

@article{schmidt-first-paper,
   author       = "P. Schmidt and M. T. Amawi and S. Pogorzalek and F. Deppe and A. Marx and R. Gross and H. Huebl",
   title        = "Sideband-resolved resonator electromechanics based on a nonlinear Josephson inductance probed on the single-photon level",
   journal      = "Commun.\ Phys.",
   year         = "2020",
   month        = "12",
   volume       = "3",
   issue        = "1",
   pages        = "1-7",
   url          = "https://www.nature.com/articles/s42005-020-00501-3",
}

@article{luschmann-mech-freq-control,
   author       = "T. Luschmann and P. Schmidt and F. Deppe and A. Marx and A. Sanchez and R. Gross and H. Huebl",
   title        = "Mechanical frequency control in inductively coupled electromechanical systems",
   journal      = "Sci.\ Rep.",
   year         = "2022",
   month        = "01",
   volume       = "12",
   issue        = "1",
   pages        = "1608",
   url          = "https://www.nature.com/articles/s41598-022-05438-x",
}

@article{Bera-singh-device-comphys2021,
   author       = "T. Bera and S. Majumder and S. K. Sahu and V. Singh",
   title        = "Large flux-mediated coupling in hybrid electromechanical system with a transmon qubit",
   journal      = "Commun Phys",
   year         = "2021",
   month        = "01",
   volume       = "4",
   issue        = "1",
   pages        = "1-7",
   url          = "https://www.nature.com/articles/s42005-020-00514-y",
}

@article{Bera-instab-natcomun2024,
   author       = "T. Bera and M. Kandpal and G. S. Agarwal and V. Singh",
   title        = "Single-photon induced instabilities in a cavity electromechanical device",
   journal      = "Nat Commun",
   year         = "2024",
   month        = "08",
   volume       = "15",
   issue        = "1",
   pages        = "7115",
   url          = "https://www.nature.com/articles/s41467-024-51499-z",
}

@article{Delattre-self-sust-destr-prres2024,
   author       = "A. Delattre and I. Golokolenov and R. Pedurand and X. Zhou and A. Fefferman and E. Collin",
   title        = "Self-sustained optomechanical state destruction triggered by the Kerr nonlinearity",
   journal      = "Phys.\ Rev.\ Res.",
   year         = "2024",
   month        = "10",
   volume       = "6",
   issue        = "4",
   pages        = "043038",
   url          = "https://link.aps.org/doi/10.1103/PhysRevResearch.6.043038",
}

@article{Doolin-nonlin-opto-stat-pra2014,
   author       = "C. Doolin and B. D. Hauer and P. H. Kim and A. J. R. MacDonald and H. Ramp and J. P. Davis",
   title        = "Nonlinear optomechanics in the stationary regime",
   journal      = "Phys.\ Rev.\ A",
   year         = "2014",
   month        = "05",
   volume       = "89",
   issue        = "5",
   pages        = "053838",
   url          = "https://link.aps.org/doi/10.1103/PhysRevA.89.053838",
}

@article{Bothner-4wave-cool-comphys2022,
   author       = "D. Bothner and I. C. Rodrigues and G. A. Steele",
   title        = "Four-wave-cooling to the single phonon level in Kerr optomechanics",
   journal      = "Commun Phys",
   year         = "2022",
   month        = "02",
   volume       = "5",
   issue        = "1",
   pages        = "1-10",
   url          = "https://www.nature.com/articles/s42005-022-00808-3",
}

@article{Majumder-cooling-qbit-mech-prr2022,
   author       = "S. Majumder and T. Bera and V. Singh",
   title        = "Prospects of cooling a mechanical resonator with a transmon qubit in c-QED setup",
   journal      = "Phys.\ Rev.\ Res.",
   year         = "2022",
   month        = "09",
   volume       = "4",
   issue        = "3",
   pages        = "033232",
   url          = "https://link.aps.org/doi/10.1103/PhysRevResearch.4.033232",
}

@article{ligo-virgo-gravitation-waves,
   author       = "LIGO Scientific Collaboration and Virgo Collaboration",
   title        = "Observation of Gravitational Waves from a Binary Black Hole Merger",
   journal      = "Phys.\ Rev.\ Lett.",
   year         = "2016",
   month        = "02",
   volume       = "116",
   issue        = "6",
   pages        = "061102",
   url          = "https://link.aps.org/doi/10.1103/PhysRevLett.116.061102",
}

@article{chan-ground-state-laser-cooling,
   author       = "J. Chan and T. P. M. Alegre and A. H. Safavi-Naeini and J. T. Hill and A. Krause and S. Gröblacher and M. Aspelmeyer and O. Painter",
   title        = "Laser cooling of a nanomechanical oscillator into its quantum ground state",
   journal      = "Nature",
   year         = "2011",
   month        = "10",
   volume       = "478",
   issue        = "7367",
   pages        = "89-92",
   url          = "https://www.nature.com/articles/nature10461",
}

@article{Regal-mech-motion-cavity-natphys2008,
   author       = "Regal, C. A. and Teufel, J. D. and Lehnert, K. W.",
   title        = "Measuring nanomechanical motion with a microwave cavity interferometer",
   journal      = "Nature Phys",
   year         = "2008",
   month        = "07",
   volume       = "4",
   issue        = "7",
   pages        = "555–560",
   url          = "https://www.nature.com/articles/nphys974",
}

@article{Barzanjeh-opto-quantum-tech,
   author       = "S. Barzanjeh and A. Xuereb and S. Gröblacher and M. Paternostro and C. A. Regal and E. M. Weig",
   title        = "Optomechanics for quantum technologies",
   journal      = "Nat.\ Phys.",
   year         = "2022",
   month        = "01",
   volume       = "18",
   issue        = "1",
   pages        = "15-24",
   url          = "https://www.nature.com/articles/s41567-021-01402-0",
}

@article{Rabl-photon-block-prl2011,
   author       = "P. Rabl",
   title        = "Photon Blockade Effect in Optomechanical Systems",
   journal      = "Phys.\ Rev.\ Lett.",
   year         = "2011",
   month        = "08",
   volume       = "107",
   issue        = "6",
   pages        = "063601",
   url          = "https://link.aps.org/doi/10.1103/PhysRevLett.107.063601",
}

@article{Nunnenkamp-single-photon-prl2011,
   author       = "A. Nunnenkamp and K. Børkje and S. M. Girvin",
   title        = "Single-Photon Optomechanics",
   journal      = "Phys. Rev. Lett.",
   year         = "2011",
   month        = "08",
   volume       = "107",
   issue        = "6",
   pages        = "063602",
   url          = "https://link.aps.org/doi/10.1103/PhysRevLett.107.063602",
}

@article{Qian-strong-coupl-prl2012,
   author       = "Jiang Qian and A. A. Clerk and K. Hammerer and F. Marquardt",
   title        = "Quantum Signatures of the Optomechanical Instability",
   journal      = "Phys.\ Rev.\ Lett.",
   year         = "2012",
   month        = "12",
   volume       = "109",
   issue        = "25",
   pages        = "253601",
   url          = "https://link.aps.org/doi/10.1103/PhysRevLett.109.253601",
}

@article{Clerk-phonon-shot-noise-prl2010,
   author       = "A. A. Clerk and F. Marquardt and J. G. E. Harris",
   title        = "Quantum Measurement of Phonon Shot Noise",
   journal      = "Phys.\ Rev.\ Lett.",
   year         = "2010",
   month        = "05",
   volume       = "104",
   issue        = "21",
   pages        = "213603",
   url          = "https://link.aps.org/doi/10.1103/PhysRevLett.104.213603",
}

@article{Kronwald-omit-nonlinear-prl2013,
   author       = "A. Kronwald and F. Marquardt",
   title        = "Optomechanically Induced Transparency in the Nonlinear Quantum Regime",
   journal      = "Phys.\ Rev.\ Lett.",
   year         = "2013",
   month        = "09",
   volume       = "111",
   issue        = "13",
   pages        = "133601",
   url          = "https://link.aps.org/doi/10.1103/PhysRevLett.111.133601",
}

@article{Hauer-QND-single-photon-pra2018,
   author       = "B. D. Hauer and A. Metelmann and J. P. Davis",
   title        = "Phonon quantum nondemolition measurements in nonlinearly coupled optomechanical cavities",
   journal      = "Phys.\ Rev.\ A",
   year         = "2018",
   month        = "10",
   volume       = "98",
   issue        = "4",
   pages        = "043804",
   url          = "https://link.aps.org/doi/10.1103/PhysRevA.98.043804",
}

@article{Liu-phonon-block-pra2010,
   author       = "Yu-xi Liu and A. Miranowicz and Y. B. Gao and J. Bajer and C. P. Sun and F. Nori",
   title        = "Qubit-induced phonon blockade as a signature of quantum behavior in nanomechanical resonators",
   journal      = "Phys.\ Rev.\ A",
   year         = "2010",
   month        = "09",
   volume       = "82",
   issue        = "3",
   pages        = "032101",
   url          = "https://link.aps.org/doi/10.1103/PhysRevA.82.032101",
}

@article{Vlastakis-100photoncat-sci2013,
   author       = "B. Vlastakis and G. Kirchmair and Z. Leghtas and S. E. Nigg and L. Frunzio and S. M. Girvin and M. Mirrahimi and M. H. Devoret and R. J. Schoelkopf",
   title        = "Deterministically Encoding Quantum Information Using 100-Photon Schrödinger Cat States",
   journal      = "Science",
   year         = "2013",
   month        = "11",
   volume       = "342",
   issue        = "6158",
   pages        = "607-610",
   url          = "https://www.science.org/doi/full/10.1126/science.1243289",
}

@article{Hofheinz-qstates-nat2009,
   author       = "M. Hofheinz and H. Wang and M. Ansmann and R. C. Bialczak and E. Lucero and M. Neeley and A. D. O’Connell and D. Sank and J. Wenner and J. M. Martinis and A. N. Cleland",
   title        = "Synthesizing arbitrary quantum states in a superconducting resonator",
   journal      = "Nature",
   year         = "2009",
   month        = "05",
   volume       = "459",
   issue        = "7246",
   pages        = "546-549",
   url          = "https://www.nature.com/articles/nature08005",
}

@article{OConnell-mech-ground-nat2010,
   author       = "A. D. O’Connell and M. Hofheinz and M. Ansmann and Radoslaw C. Bialczak and M. Lenander and Erik Lucero and M. Neeley and D. Sank and H. Wang and M. Weides and J. Wenner and John M. Martinis and A. N. Cleland",
   title        = "Quantum ground state and single-phonon control of a mechanical resonator",
   journal      = "Nature",
   year         = "2010",
   month        = "04",
   volume       = "464",
   issue        = "7289",
   pages        = "697-703",
   url          = "https://www.nature.com/articles/nature08967",
}

@article{Teufel-mech-ground-nat2011,
   author       = "J. D. Teufel and T. Donner and Dale Li and J. W. Harlow and M. S. Allman and K. Cicak and A. J. Sirois and J. D. Whittaker and K. W. Lehnert and R. W. Simmonds",
   title        = "Sideband cooling of micromechanical motion to the quantum ground state",
   journal      = "Nature",
   year         = "2011",
   month        = "07",
   volume       = "475",
   issue        = "7356",
   pages        = "359-363",
   url          = "https://www.nature.com/articles/nature10261",
}

@article{Teufel-strong-coupl-nat2011,
   author       = "Teufel, J. D. and Li, D. and Allman, M. S. and Cicak, K. and Sirois, A. J. and Whittaker, J. D. and Simmonds, R. W.",
   title        = "Circuit cavity electromechanics in the strong-coupling regime",
   journal      = "Nature",
   year         = "2011",
   month        = "03",
   volume       = "471",
   issue        = "7337",
   pages        = "204-208",
   url          = "https://www.nature.com/articles/nature09898",
}

@article{Singh-omit-graphene-steele-natnano2014,
   author       = "Singh, V. and Bosman, S. J. and Schneider, B. H. and Blanter, Y. M. and Castellanos-Gomez, A. and Steele, G. A.",
   title        = "Optomechanical coupling between a multilayer graphene mechanical resonator and a superconducting microwave cavity",
   journal      = "Nature Nanotech",
   year         = "2014",
   month        = "10",
   volume       = "9",
   issue        = "10",
   pages        = "820-824",
   url          = "https://www.nature.com/articles/nnano.2014.168",
}

@article{Weber-g0-t-dep-natcomun2016,
   author       = "Weber, P. and Güttinger, J. and Noury, A. and Vergara-Cruz, J. and Bachtold, A.",
   title        = "Force sensitivity of multilayer graphene optomechanical devices",
   journal      = "Nat Commun",
   year         = "2016",
   month        = "08",
   volume       = "7",
   issue        = "1",
   pages        = "12496",
   url          = "https://www.nature.com/articles/ncomms12496",
}

@article{bruno-nc-sc-res-applphyslett2015,
   author       = "Bruno, A. and De Lange, G. and Asaad, S. and Van Der Enden, K. L. and Langford, N. K. and DiCarlo, L.",
   title        = "Reducing intrinsic loss in superconducting resonators by surface treatment and deep etching of silicon substrates",
   journal      = "Appl.\ Phys.\ Lett.",
   year         = "2015",
   month        = "05",
   volume       = "106",
   issue        = "18",
   pages        = "182601",
   url          = "https://pubs.aip.org/apl/article/106/18/182601/27784/Reducing-intrinsic-loss-in-superconducting",
}

@article{schmidt-n-ph-calib-applphyslett2018,
   author       = "Schmidt, P. and Schwienbacher, D. and Pernpeintner, M. and Wulschner, F. and Deppe, F. and Marx, A. and Gross, R. and Huebl, H.",
   title        = "Ultrawide-range photon number calibration using a hybrid system combining nano-electromechanics and superconducting circuit quantum electrodynamics",
   journal      = "Appl.\ Phys.\ Lett.",
   year         = "2018",
   month        = "10",
   volume       = "113",
   issue        = "15",
   pages        = "152601",
   url          = "https://pubs.aip.org/apl/article/113/15/152601/35024/Ultrawide-range-photon-number-calibration-using-a",
}

@article{Wollmann-mech-squeez-sci2015,
   author       = "E. E. Wollmann and C. U. Lei and A. J. Weinstein and J. Suh and A. Kronwald and F. Marquardt and A. A. Clerk and K. C. Schwab",
   title        = "Quantum squeezing of motion in a mechanical resonator",
   journal      = "Science",
   year         = "2015",
   month        = "08",
   volume       = "349",
   issue        = "6251",
   pages        = "952-955",
   url          = "https://www.science.org/doi/10.1126/science.aac5138",
}

@article{Pirkkalainen-mech-squeez-prl2015,
  title = "Squeezing of Quantum Noise of Motion in a Micromechanical Resonator",
  author = "J.-M. Pirkkalainen and E. Damskägg and M. Brandt and F. Massel and M. A. Sillanpää",
  journal = "Phys. Rev. Lett.",
  year = "2015",
  month = "12",
  volume = "115",
  issue = "24",
  pages = "243601",
  url = "https://link.aps.org/doi/10.1103/PhysRevLett.115.243601",
}

@article{Barzanjeh-entangle-mech-nature2019,
   author       = "S. Barzanjeh and E. S. Redchenko and M. Peruzzo and M. Wulf and D. P. Lewis and G. Arnold and J. M. Fink",
   title        = "Stationary entangled radiation from micromechanical motion",
   journal      = "Nature",
   year         = "2019",
   month        = "06",
   volume       = "570",
   issue        = "7762",
   pages        = "480-483",
   url          = "https://www.nature.com/articles/s41586-019-1320-2",
}

@article{Lecocq-nondem-massive-prx2015,
   author       = "F. Lecocq and J. B.Clark and R. W. Simmonds and J. Aumentado and J. D. Teufel",
   title        = "Quantum Nondemolition Measurement of a Nonclassical State of a Massive Object",
   journal      = "Phys.\ Rev.\ X",
   year         = "2015",
   month        = "12",
   volume       = "5",
   issue        = "4",
   pages        = "041037",
   url          = "https://link.aps.org/doi/10.1103/PhysRevX.5.041037",
}

@article{Barzanjeh-on-chip-circulator-natcom2017,
   author       = "S. Barzanjeh and M. Wulf and M. Peruzzo and M. Kalaee and P. B. Dieterle and O. Painter and J. M. Fink",
   title        = "Mechanical on-chip microwave circulator",
   journal      = "Nat Commun",
   year         = "2017",
   month        = "10",
   volume       = "8",
   issue        = "1",
   pages        = "953",
   url          = "https://www.nature.com/articles/s41467-017-01304-x",
}

@article{carmon-selfosc-2007,
  title = {Temporal Behavior of Radiation-Pressure-Induced Vibrations of an Optical Microcavity Phonon Mode},
  author = {Carmon, Tal and Rokhsari, Hossein and Yang, Lan and Kippenberg, Tobias J. and Vahala, Kerry J.},
  journal = {Phys. Rev. Lett.},
  volume = {94},
  issue = {22},
  pages = {223902},
  numpages = {4},
  year = {2005},
  month = {Jun},
  publisher = {American Physical Society},
  doi = {10.1103/PhysRevLett.94.223902},
  url = {https://link.aps.org/doi/10.1103/PhysRevLett.94.223902}
}

@article{probst-circle-fit,
   author       = "S. Probst and F. B. Song and P. A. Bushev and A. V. Ustinov and M. Weides",
   title        = "Efficient and robust analysis of complex scattering data under noise in microwave resonators",
   journal      = "Rev.\ Sci.\ Instrum.",
   year         = "2015",
   month        = "02",
   volume       = "86",
   issue        = "2",
   pages        = "024706",
   url          = "https://pubs.aip.org/rsi/article/86/2/024706/360955/Efficient-and-robust-analysis-of-complex",
}

@article{deeg-lukas-pra012025-backaction-bistable,
   author       = "L. F. Deeg and D. Zoepfl and N. Diaz-Naufal and M. L. Juan and A. Metelmann and G. Kirchmair",
   title        = "Optomechanical backaction in the bistable regime",
   journal      = "Phys.\ Rev.\ Appl.",
   year         = "2025",
   month        = "01",
   volume       = "23",
   issue        = "1",
   pages        = "014082",
   url          = "https://link.aps.org/doi/10.1103/PhysRevApplied.23.014082",
}

@article{nico-nonlinear-cooling,
   author       = "Diaz-Naufal, N. and Deeg, L. and Zoepfl, D. and Schneider, C. M. F. and Juan, M. L. and Kirchmair, G. and Metelmann, A.",
   title        = "Kerr-enhanced optomechanical cooling in the unresolved-sideband regime",
   journal      = "Phys.\ Rev.\ A",
   year         = "2025",
   month        = "05",
   volume       = "111",
   issue        = "5",
   pages        = "053505",
   url          = "https://link.aps.org/doi/10.1103/PhysRevA.111.053505",
}

@article{zoepfl-nonlinear-cooling,
   author       = "D. Zoepfl and M. L. Juan and N. Diaz-Naufal and C. M. F. Schneider and L. F. Deeg and A. Sharafiev and A. Metelmann and G. Kirchmair",
   title        = "Kerr Enhanced Backaction Cooling in Magnetomechanics",
   journal      = "Phys.\ Rev.\ Lett.",
   year         = "2023",
   month        = "01",
   volume       = "130",
   issue        = "3",
   pages        = "033601",
   url          = "https://link.aps.org/doi/10.1103/PhysRevLett.130.033601",
}

@article{he-fast-cat-kerr-tuning,
   author       = "X. L. He and Y. Lu and D. Q. Bao and H. Xue and W. B. Jiang and Z. Wang and A. F. Roudsari and P. Delsing and J. S. Tsai and Z. R. Lin",
   title        = "Fast generation of Schrödinger cat states using a Kerr-tunable superconducting resonator",
   journal      = "Nat.\ Commun.",
   year         = "2023",
   month        = "10",
   volume       = "14",
   issue        = "1",
   pages        = "6358",
   url          = "https://www.nature.com/articles/s41467-023-42057-0",
}

@article{puri-quantum-state-kerr-two-photon,
   author       = "S. Puri and S. Boutin and A. Blais",
   title        = "Engineering the quantum states of light in a Kerr-nonlinear resonator by two-photon driving",
   journal      = "npj Quantum Inf.",
   year         = "2017",
   month        = "04",
   volume       = "3",
   issue        = "1",
   pages        = "1-7",
   url          = "https://www.nature.com/articles/s41534-017-0019-1",
}

@Book{hopf-bifurcation,
   author    = "L.N. Howard and N. Kopell",
   title     = "A Translation of Hopf’s Original Paper",
   volume    = "19",
   booktitle = "The Hopf Bifurcation and Its Applications",
   publisher = "Springer",
   address   = "New York",
   year      = "1976",
   url       = "https://link.springer.com/book/10.1007/978-1-4612-6374-6#back-to-top",
}

@article{gorodetksy-g0-determination,
   author       = "M. L. Gorodetksy and A. Schliesser and G. Anetsberger and S. Deleglise and T. J. Kippenberg",
   title        = "Determination of the vacuum optomechanical coupling rate using frequency noise calibration",
   journal      = "Opt.\ Express",
   year         = "2010",
   month        = "10",
   volume       = "18",
   issue        = "22",
   pages        = "23236-23246",
   url          = "https://opg.optica.org/oe/abstract.cfm?uri=oe-18-22-23236",
}

@article{Weis-OMIT,
   author       = "S. Weis and R. Rivière and S. Deléglise and E. Gavartin and O. Arcizet and A. Schliesser and T. J. Kippenberg",
   title        = "Optomechanically Induced Transparency",
   journal      = "Science",
   year         = "2010",
   month        = "12",
   volume       = "330",
   issue        = "6010",
   pages        = "1520-1523",
   url          = "https://www.science.org/doi/10.1126/science.1195596",
}

@article{Hauer-cool-to-cat-2023,
   author       = "B. D. Hauer and J. Combes and J. D. Teufel",
   title        = "Nonlinear Sideband Cooling to a Cat State of Motion",
   journal      = "Phys.\ Rev.\ Lett.",
   year         = "2023",
   month        = "05",
   volume       = "130",
   issue        = "21",
   pages        = "213604",
   url          = "https://link.aps.org/doi/10.1103/PhysRevLett.130.213604",
}

@article{Zhou-nanoelectromech,
   author       = "X. Zhou and F. Hocke and A. Schliesser and A. Marx and H. Huebl and R. Gross and T. J. Kippenberg",
   title        = "Slowing, advancing and switching of microwave signals using circuit nanoelectromechanics",
   journal      = "Nature Phys.",
   year         = "2013",
   month        = "03",
   volume       = "9",
   issue        = "3",
   pages        = "179-184",
   url          = "https://www.nature.com/articles/nphys2527",
}

@article{aspelmeyer-cavity-optomechanics,
   author       = "M. Aspelmeyer and T. J. Kippenberg and F. Marquardt",
   title        = "Cavity optomechanics",
   journal      = "Rev.\ Mod.\ Phys.",
   year         = "2014",
   month        = "12",
   volume       = "86",
   issue        = "4",
   pages        = "1391-1452",
   url          = "https://link.aps.org/doi/10.1103/RevModPhys.86.1391",
}

@article{li-cav-optom-sens-nanophoto2021,
   author       = "B. Li and L. Ou and Y. Lei and Y. Liu",
   title        = "Cavity optomechanical sensing",
   journal      = "Nanophotonics",
   year         = "2021",
   month        = "09",
   volume       = "10",
   issue        = "11",
   pages        = "2799-2832",
   url          = "https://www.degruyter.com/document/doi/10.1515/nanoph-2021-0256/html",
}

@article{Metcalfe-apl-cav-optomech-aprev2014,
   author       = "M. Metcalfe",
   title        = "Applications of cavity optomechanics",
   journal      = "Appl.\ Phys.\ Rev.",
   year         = "2014",
   month        = "09",
   volume       = "1",
   issue        = "3",
   pages        = "031105",
   url          = "https://doi.org/10.1063/1.4896029",
}

@article{rodrigues-first-nanoelectromech-device,
   author       = "I. C. Rodrigues and D. Bothner and G. A. Steele",
   title        = "Coupling microwave photons to a mechanical resonator using quantum interferences",
   journal      = "Nat Commun",
   year         = "2019",
   month        = "11",
   volume       = "10",
   issue        = "1",
   pages        = "5359",
   url          = "https://www.nature.com/articles/s41467-019-12964-2",
}

@article{Kuang-phonon-laser-optomech-natphys2023,
   author       = "Kuang, T. and Huang, R. and Xiong, W. and Zuo, Y. and Han, X. and Nori, F. and Qiu, C. and Luo, H. and Jing, H. and Xiao, G.",
   title        = "Nonlinear multi-frequency phonon lasers with active levitated optomechanics",
   journal      = "Nat.\ Phys.",
   year         = "2023",
   month        = "03",
   volume       = "19",
   issue        = "3",
   pages        = "414-419",
   url          = "https://www.nature.com/articles/s41567-022-01857-9",
}

@article{Huber-squeez-mech-mode-prx2020,
   author       = "Huber, J. S. and Rastelli, G. and Seitner, M. J. and Kölbl, J. and Belzig, W. and Dykman, M. I. and Weig, E. M.",
   title        = "Spectral Evidence of Squeezing of a Weakly Damped Driven Nanomechanical Mode",
   journal      = "Phys.\ Rev.\ X",
   year         = "2020",
   month        = "06",
   volume       = "10",
   issue        = "2",
   pages        = "021066",
   url          = "https://link.aps.org/doi/10.1103/PhysRevX.10.021066",
}

@article{Ochs-freq-comb-mech-mode-prx2022,
   author       = "Ochs, J. S. and Boneß, D. K. J. and Rastelli, G. and Seitner, M. and Belzig, W. and Dykman, M. I. and Weig, E. M.",
   title        = "Frequency Comb from a Single Driven Nonlinear Nanomechanical Mode",
   journal      = "Phys.\ Rev.\ X",
   year         = "2022",
   month        = "11",
   volume       = "12",
   issue        = "4",
   pages        = "041019",
   url          = "https://link.aps.org/doi/10.1103/PhysRevX.12.041019",
}

@article{Rieger-fano-mw-res-prapp2023,
   author       = "Rieger, D. and Günzler, S. and Spiecker, M. and Nambisan, A. and Wernsdorfer, W. and Pop, I.M.",
   title        = "Fano Interference in Microwave Resonator Measurements",
   journal      = "Phys.\ Rev.\ Appl.",
   year         = "2023",
   month        = "07",
   volume       = "20",
   issue        = "1",
   pages        = "014059",
   url          = "https://link.aps.org/doi/10.1103/PhysRevApplied.20.014059",
}

@article{Khalil-s21-notch-res-japplphys2012,
   author       = "Khalil, M. S. and Stoutimore, M. J. A. and Wellstood, F. C. and Osborn, K. D.",
   title        = "An analysis method for asymmetric resonator transmission applied to superconducting devices",
   journal      = "J.\ Appl.\ Phys.",
   year         = "2012",
   month        = "03",
   volume       = "111",
   issue        = "5",
   pages        = "054510",
   url          = "https://doi.org/10.1063/1.3692073",
}

@article{Brennecke-bose-einstein-science2008,
   author       = "F. Brennecke and S. Ritter and T. Donner and T. Esslinger",
   title        = "Cavity Optomechanics with a Bose-Einstein Condensate",
   journal      = "Science",
   year         = "2008",
   month        = "10",
   volume       = "322",
   issue        = "5899",
   pages        = "235-238",
   url          = "https://www.science.org/doi/full/10.1126/science.1163218",
}

@article{Armour2012,
title = {Quantum dynamics of a mechanical resonator driven by a cavity},
journal = {Comptes Rendus Physique},
volume = {13},
number = {5},
pages = {440-453},
year = {2012},
note = {Advances in nano-electromechanical systems},
issn = {1631-0705},
doi = {https://doi.org/10.1016/j.crhy.2012.03.006},
url = {https://www.sciencedirect.com/science/article/pii/S1631070512000382},
author = {Andrew D. Armour and Denzil A. Rodrigues}
}

@article{rodrigues-2010,
  title = {Amplitude Noise Suppression in Cavity-Driven Oscillations of a Mechanical Resonator},
  author = {Rodrigues, D. A. and Armour, A. D.},
  journal = {Phys. Rev. Lett.},
  volume = {104},
  issue = {5},
  pages = {053601},
  numpages = {4},
  year = {2010},
  month = {Feb},
  publisher = {American Physical Society},
  doi = {10.1103/PhysRevLett.104.053601},
  url = {https://link.aps.org/doi/10.1103/PhysRevLett.104.053601}
}

@article{aldana2013,
  title = {Equivalence between an optomechanical system and a Kerr medium},
  author = {Aldana, Samuel and Bruder, Christoph and Nunnenkamp, Andreas},
  journal = {Phys. Rev. A},
  volume = {88},
  issue = {4},
  pages = {043826},
  numpages = {10},
  year = {2013},
  month = {Oct},
  publisher = {American Physical Society},
  doi = {10.1103/PhysRevA.88.043826},
  url = {https://link.aps.org/doi/10.1103/PhysRevA.88.043826}
}

@article{laflamme2011,
  title = {Quantum-limited amplification with a nonlinear cavity detector},
  author = {Laflamme, C. and Clerk, A. A.},
  journal = {Phys. Rev. A},
  volume = {83},
  issue = {3},
  pages = {033803},
  numpages = {13},
  year = {2011},
  month = {Mar},
  publisher = {American Physical Society},
  doi = {10.1103/PhysRevA.83.033803},
  url = {https://link.aps.org/doi/10.1103/PhysRevA.83.033803}
}

@article{yurke-cat-state-1986,
  title = {Generating quantum mechanical superpositions of macroscopically distinguishable states via amplitude dispersion},
  author = {Yurke, B. and Stoler, D.},
  journal = {Phys. Rev. Lett.},
  volume = {57},
  issue = {1},
  pages = {13--16},
  numpages = {0},
  year = {1986},
  month = {Jul},
  publisher = {American Physical Society},
  doi = {10.1103/PhysRevLett.57.13},
  url = {https://link.aps.org/doi/10.1103/PhysRevLett.57.13}
}

@article{Wang_2021-hangermode,
doi = {10.1088/2058-9565/ac070e},
url = {https://dx.doi.org/10.1088/2058-9565/ac070e},
year = {2021},
month = {jun},
publisher = {IOP Publishing},
volume = {6},
number = {3},
pages = {035015},
author = {Wang, Haozhi and Singh, S and McRae, C R H and Bardin, J C and Lin, S-X and Messaoudi, N and Castelli, A R and Rosen, Y J and Holland, E T and Pappas, D P and Mutus, J Y},
title = {Cryogenic single-port calibration for superconducting microwave resonator measurements},
journal = {Quantum Science and Technology}
}

@book{dykman-nonlinear-oscillators-2012-book,
    author = {Dykman, Mark},
    title = {Fluctuating Nonlinear Oscillators: From Nanomechanics to Quantum Superconducting Circuits},
    publisher = {Oxford University Press},
    year = {2012},
    month = {07},
    isbn = {9780199691388},
    doi = {10.1093/acprof:oso/9780199691388.001.0001},
    url = {https://doi.org/10.1093/acprof:oso/9780199691388.001.0001},
}

@article{Pdnation2013,
  title = {Nonclassical mechanical states in an optomechanical micromaser analog},
  author = {Nation, P. D.},
  journal = {Phys. Rev. A},
  volume = {88},
  issue = {5},
  pages = {053828},
  numpages = {7},
  year = {2013},
  month = {Nov},
  publisher = {American Physical Society},
  doi = {10.1103/PhysRevA.88.053828},
  url = {https://link.aps.org/doi/10.1103/PhysRevA.88.053828}
}

@article{srdas2023,
  title = {Instabilities near Ultrastrong Coupling in a Microwave Optomechanical Cavity},
  author = {Das, Soumya Ranjan and Majumder, Sourav and Sahu, Sudhir Kumar and Singhal, Ujjawal and Bera, Tanmoy and Singh, Vibhor},
  journal = {Phys. Rev. Lett.},
  volume = {131},
  issue = {6},
  pages = {067001},
  numpages = {6},
  year = {2023},
  month = {Aug},
  publisher = {American Physical Society},
  doi = {10.1103/PhysRevLett.131.067001},
  url = {https://link.aps.org/doi/10.1103/PhysRevLett.131.067001}
}

@article{gardiner1985,
  title = {Input and output in damped quantum systems: Quantum stochastic differential equations and the master equation},
  author = {Gardiner, C. W. and Collett, M. J.},
  journal = {Phys. Rev. A},
  volume = {31},
  issue = {6},
  pages = {3761--3774},
  numpages = {0},
  year = {1985},
  month = {Jun},
  publisher = {American Physical Society},
  doi = {10.1103/PhysRevA.31.3761},
  url = {https://link.aps.org/doi/10.1103/PhysRevA.31.3761}
}

@article{nation2008,
  title = {Quantum analysis of a nonlinear microwave cavity-embedded dc SQUID displacement detector},
  author = {Nation, P. D. and Blencowe, M. P. and Buks, E.},
  journal = {Phys. Rev. B},
  volume = {78},
  issue = {10},
  pages = {104516},
  numpages = {17},
  year = {2008},
  month = {Sep},
  publisher = {American Physical Society},
  doi = {10.1103/PhysRevB.78.104516},
  url = {https://link.aps.org/doi/10.1103/PhysRevB.78.104516}
}

@article{Ludwig2008,
doi = {10.1088/1367-2630/10/9/095013},
url = {https://dx.doi.org/10.1088/1367-2630/10/9/095013},
year = {2008},
month = {sep},
publisher = {},
volume = {10},
number = {9},
pages = {095013},
author = {Ludwig, Max and Kubala, Björn and Marquardt, Florian},
title = {The optomechanical instability in the quantum regime},
journal = {New Journal of Physics}
}

@article{marquardt2006,
  title = {Dynamical Multistability Induced by Radiation Pressure in High-Finesse Micromechanical Optical Cavities},
  author = {Marquardt, Florian and Harris, J. G. E. and Girvin, S. M.},
  journal = {Phys. Rev. Lett.},
  volume = {96},
  issue = {10},
  pages = {103901},
  numpages = {4},
  year = {2006},
  month = {Mar},
  publisher = {American Physical Society},
  doi = {10.1103/PhysRevLett.96.103901},
  url = {https://link.aps.org/doi/10.1103/PhysRevLett.96.103901}
}

@article{krause2015,
  title = {Nonlinear Radiation Pressure Dynamics in an Optomechanical Crystal},
  author = {Krause, Alex G. and Hill, Jeff T. and Ludwig, Max and Safavi-Naeini, Amir H. and Chan, Jasper and Marquardt, Florian and Painter, Oskar},
  journal = {Phys. Rev. Lett.},
  volume = {115},
  issue = {23},
  pages = {233601},
  numpages = {5},
  year = {2015},
  month = {Dec},
  publisher = {American Physical Society},
  doi = {10.1103/PhysRevLett.115.233601},
  url = {https://link.aps.org/doi/10.1103/PhysRevLett.115.233601}
}

@article{roque2020,
    author = {Thales Figueiredo Roque},
    title = {Nonlinear dynamics of weakly dissipative optomechanical systems},
    journal ={ New J. Phys},
    volume={22},
    pages={013049},
    year = {2020},
    url = {https://dx.doi.org/10.1088/1367-2630/ab6522}
}

@article{reviewdegen2017,
  title = {Quantum sensing},
  author = {Degen, C. L. and Reinhard, F. and Cappellaro, P.},
  journal = {Rev. Mod. Phys.},
  volume = {89},
  issue = {3},
  pages = {035002},
  numpages = {39},
  year = {2017},
  month = {Jul},
  publisher = {American Physical Society},
  doi = {10.1103/RevModPhys.89.035002},
  url = {https://link.aps.org/doi/10.1103/RevModPhys.89.035002}
}

@article{amitai2017,
  title = {Synchronization of an optomechanical system to an external drive},
  author = {Amitai, Ehud and L\"orch, Niels and Nunnenkamp, Andreas and Walter, Stefan and Bruder, Christoph},
  journal = {Phys. Rev. A},
  volume = {95},
  issue = {5},
  pages = {053858},
  numpages = {9},
  year = {2017},
  month = {May},
  publisher = {American Physical Society},
  doi = {10.1103/PhysRevA.95.053858},
  url = {https://link.aps.org/doi/10.1103/PhysRevA.95.053858}
}

@article{jensen1998,
  title = {Synchronization of randomly driven nonlinear oscillators},
  author = {Jensen, R. V.},
  journal = {Phys. Rev. E},
  volume = {58},
  issue = {6},
  pages = {R6907--R6910},
  numpages = {0},
  year = {1998},
  month = {Dec},
  publisher = {American Physical Society},
  doi = {10.1103/PhysRevE.58.R6907},
  url = {https://link.aps.org/doi/10.1103/PhysRevE.58.R6907}
}

@article{bakemeier2015,
  title = {Route to Chaos in Optomechanics},
  author = {Bakemeier, L. and Alvermann, A. and Fehske, H.},
  journal = {Phys. Rev. Lett.},
  volume = {114},
  issue = {1},
  pages = {013601},
  numpages = {5},
  year = {2015},
  month = {Jan},
  publisher = {American Physical Society},
  doi = {10.1103/PhysRevLett.114.013601},
  url = {https://link.aps.org/doi/10.1103/PhysRevLett.114.013601}
}

@article{lu2015,
  title = {$\mathcal{P}\mathcal{T}$-Symmetry-Breaking Chaos in Optomechanics},
  author = {L\"u, Xin-You and Jing, Hui and Ma, Jin-Yong and Wu, Ying},
  journal = {Phys. Rev. Lett.},
  volume = {114},
  issue = {25},
  pages = {253601},
  numpages = {6},
  year = {2015},
  month = {Jun},
  publisher = {American Physical Society},
  doi = {10.1103/PhysRevLett.114.253601},
  url = {https://link.aps.org/doi/10.1103/PhysRevLett.114.253601}
}

@book{Strogatz-nonlin-dyn-chaos,
    author = {S. Strogatz},
    title = {Nonlinear dynamics and chaos: with applications to physics, biology, chemistry, and engineering},
    publisher = {CRC Press},
    year = {2019}
}

@phdthesis{ludwig2013,
    author = "Max Ludwig",
    title = "Collective quantum effects in optomechanical systems",
    school= "Friedrich-Alexander-Universität Erlangen-Nürnberg (FAU)",
    year = "2013",
    url = {https://open.fau.de/items/c8385804-0d44-4191-93dc-2a86ac241c24/full}
}

\clearpage

\onecolumngrid

\section{Device Fabrication} \label{SI:sec:fab-layout}

The device is structured on a high-resistivity silicon wafer ($R > \SI{10}{\kohm\centi\meter}$), from which a $6$ by $\SI{10}{\milli\meter}$ chip is diced. The entire layout is defined in a single step using electron beam lithography. Subsequently, a $\SI{40}{\nano\meter}$ thick aluminum film is evaporated in a shadow-angle configuration, followed by oxidation, and then a deposition of a secondary $\SI{70}{\nano\meter}$ thick aluminum film. The structure is then transferred onto the substrate through a lift-off process. To obtain tensile stressed aluminum, the entire chip is annealed for $\SI{30}{\min}$ at $\SI{350}{\degreeCelsius}$ under atmospheric pressure. Following this, selective removal of the silicon beneath the aluminum is performed using reactive-ion etching (RIE), leaving two suspended and mechanically compliant nanostrings within the dc-SQUID. For further details on the fabrication process and detailed scanning electron microscopy (SEM) images, we refer to Ref.~\cite{schmidt-first-paper}.

\section{Cryogenic Wiring} \label{SI:sec:mw-setup}

Since all experiments presented here are conducted at millikelvin temperatures in a commercial dilution refrigerator, cryogenic microwave wiring is required to access the sample. The input microwave line is equipped with a total of $\SI{50}{\decibel}$ of attenuation before reaching the device under test (DUT) to suppress thermal noise and unwanted reflections. The output signal is amplified by a high-electron-mobility transistor (HEMT), which is mounted at the $\SI{4}{\kelvin}$ stage. To prevent back reflections from the amplifier, three microwave circulators are placed between the DUT and the HEMT. In addition to the microwave wiring, a DC current line allows to control the current through a superconducting coil mounted outside of the DUT package. This allows for the adjustment of the applied out-of-plane magnetic field $B_\mathrm{oop}$, which tunes the resonance frequency of the microwave cavity by modifying the effective Josephson inductance of the dc-SQUID. The large in-plane magnetic field $B_\mathrm{ip}$ is generated by a commercial 3D vector magnet, in the center of which the DUT with the small coil is placed.

\section{Device Parameter Determination} \label{SI:sec:device-params}

\begin{figure}[t]
    \centering
    \includegraphics{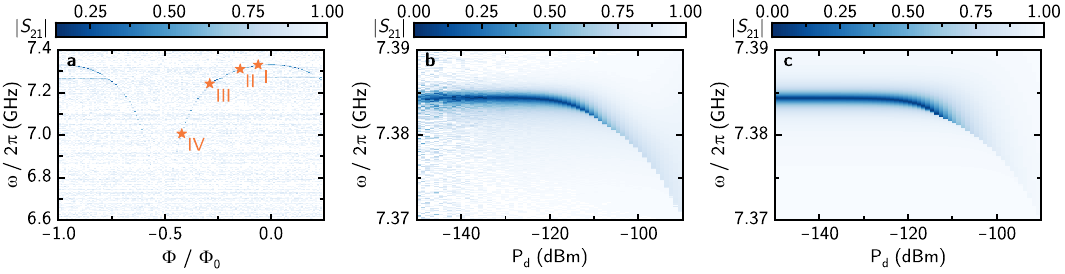}
    \caption{\textbf{Cavity parameter determination.} \textbf{a}, Flux tuning curve measured at $B_\mathrm{ip} = \SI{30}{\milli\tesla}$. The four selected working points corresponding to parameter sets I to IV, which are used in the presented measurements, are marked with orange stars. Besides the flux-tunable microwave resonator, the resonance of a fixed frequency resonator coupled to the same feedline can be seen at around $\SI{7.274}{\giga\hertz}$. We avoid this frequency range to ensure that this parasitic resonance does not affect the results. \textbf{b}, Power sweep performed at the top of the tuning curve, where $\omega_\mathrm{c}$ is maximal and $g_0 \approx 0$. In this regime, only the Kerr-induced shift of the cavity resonance towards lower frequencies is observed with increasing power. No additional absorption dips appear on the blue sideband of the cavity, as the optomechanical interaction is effectively turned off. We fit the power-induced frequency shift of the cavity resonance to obtain the Kerr nonlinearity $\mathcal{K}$. Note that this measurement was conducted at $B_\mathrm{ip} = \SI{10}{\milli\tesla}$. \textbf{c}, Analytical calculation of the power-dependent scattering response based on the fit results from \textbf{b}. The apparent excellent agreement between theory and experiment validates the value of the Kerr nonlinearity $\mathcal{K}$ obtained by the fit.}    
    \label{SI:fig:coil-kerr-sweep}
\end{figure}

To first determine the frequency range we can access with the flux-tunable microwave cavity, we set the in-plane magnetic field to $B_\mathrm{ip} = \SI{30}{\milli\tesla}$. The flux-tuning is then determined by varying the current applied to the small superconducting coil mounted to the DUT package, which sets the out-of-plane magnetic field. For each current, we perform a broad frequency sweep of the transmission parameter $S_\mathrm{21}$ from $6.6$ to $\SI{7.4}{\giga\hertz}$ using a vector network analyzer (VNA). We then extract the maximum frequency of the tuning curve and the periodicity of the tuning behavior to translate the current applied to the coil into a flux bias. The resulting flux tuning curve is shown in Fig.~\ref{SI:fig:coil-kerr-sweep}a. We now select four working points corresponding to the parameter sets I to IV, as indicated by the orange stars, for further study. At each of these points, we determine the cavity and mechanical parameters as well as the single-photon coupling rate.

\subsection{Cavity Parameters} \label{SI:sec:cavity-params}

Before any other system parameters can be determined, it is essential to first characterize the cavity. This is done using a modified circle fitting algorithm, based on the methods presented in Refs.~\cite{probst-circle-fit,deeg-lukas-pra012025-backaction-bistable}, which also allows for the extraction of the Kerr nonlinearity $\mathcal{K}$. The following section describes this procedure in greater detail.

Figure \ref{SI:fig:coil-kerr-sweep}b illustrates the scattering response of the cavity with increasing probe power. For this measurement, the cavity is operated at an in-plane magnetic field of $B_\mathrm{ip} = \SI{10}{\milli\tesla}$ and at its maximum frequency, where $\partial \omega_\mathrm{c} / \partial \Phi \approx 0$ and thus also $g_0/\left(2\pi\right) \approx 0$. Hence, we do not observe any additional absorption dips on the blue sideband here. To determine the cavity parameters, the lowest power frequency response is first fitted using the established circle fitting method, where the scattering response of a notch-type resonator is given by \cite{Khalil-s21-notch-res-japplphys2012,probst-circle-fit,Rieger-fano-mw-res-prapp2023}
\begin{equation}
    S_{21}^{\text{notch}}\left( \omega \right) = ae^{i\alpha} e^{- i \omega \tau} 
    \left[1 - \frac{\left(Q / |Q_\mathrm{ext}|\right) e^{i\phi}}{1 + 2i Q \left( \omega / \omega_\mathrm{c} - 1 \right)} \right].
    \label{SI:eq:S21-probst}
\end{equation}
Here, $\omega$ denotes the probe frequency, $a$ the attenuation/gain, $\alpha$ a phase offset, $\tau$ the electronic delay, $\phi$ the impedance mismatch, and $Q = Q_\mathrm{int} + Q_\mathrm{ext}$ the loaded quality factor with $Q_\mathrm{ext} = \omega_\mathrm{c} / \kappa_\mathrm{ext}$ and $Q_\mathrm{int} = \omega_\mathrm{c} / \kappa_\mathrm{int}$ being the external and internal quality factors, respectively. The second part of the equation in the square brackets describes an ideal notch-type resonator, whereas the first part accounts for the environment. From the low-power fit, we extract the resonance frequency $\omega_\mathrm{c}/\left(2\pi\right) = \SI{7.384}{\giga\hertz}$ and the loss rates $\kappa_\mathrm{int} / \left(2\pi\right) = \SI{0.13}{\mega\hertz}$ and $\kappa_\mathrm{ext} / \left(2\pi\right) = \SI{1.93}{\mega\hertz}$ for the data shown in Fig.~\ref{SI:fig:coil-kerr-sweep}b.

For higher powers, the resonance frequency $\omega_\mathrm{c}$ remains fixed during the following fitting procedure. Instead, the circle-fitting routine is adapted to incorporate the Kerr nonlinearity by iteratively optimizing $\mathcal{K}$ such that it correctly describes the induced frequency shift \cite{deeg-lukas-pra012025-backaction-bistable}
\begin{gather}
    \omega_\mathrm{c} \rightarrow \omega_\mathrm{c} - \mathcal{K} \bar{n}_{\mathrm{c}}, \label{SI:eq:omega_c-kerr} \\
    \bar{n}_{\mathrm{c}} \left[ \left( \omega_\mathrm{c} - \mathcal{K}\bar{n}_{\mathrm{c}} - \omega \right)^2 + \left( \frac{\omega_\mathrm{c}}{2 Q} \right)^2 \right] = \frac{\omega_\mathrm{c}}{2 \left| Q_\mathrm{c} \right|} \frac{P_\mathrm{d}}{\hbar \omega}, \label{SI:eq:n_c-kerr}
\end{gather}
where $\bar{n}_{\mathrm{c}}$ denotes the number of photons within the cavity, while $P_\mathrm{d}$ represents the power of the probe signal at its specified probe frequency $\omega$. We repeat this for all measured powers until the obtained value for $\mathcal{K}$ is stable over several powers. For the data shown in Fig.~\ref{SI:fig:coil-kerr-sweep}, we find $\mathcal{K}/\left(2\pi\right) = \SI{14}{\kilo\hertz}$. We refer to Ref.~\cite{deeg-lukas-pra012025-backaction-bistable} for more details on the extraction of the Kerr nonlinearity. Note that a photon number calibration is necessary since precise knowledge of $P_\mathrm{d}$ is required. 

With this, we have determined all cavity characteristics needed for calculating the power-dependent scattering response via Eqs.~\ref{SI:eq:S21-probst}, \ref{SI:eq:omega_c-kerr}, and \ref{SI:eq:n_c-kerr}, thereby validating the fit results. The cavity response calculated in this way, based on the fit results, is shown in Fig.~\ref{SI:fig:coil-kerr-sweep}c. It shows excellent agreement with the corresponding experimental data in Fig.~\ref{SI:fig:coil-kerr-sweep}b.

\subsection{Mechanical Parameters}  \label{SI:sec:mech-param}

\begin{figure}[t]
    \centering
    \includegraphics{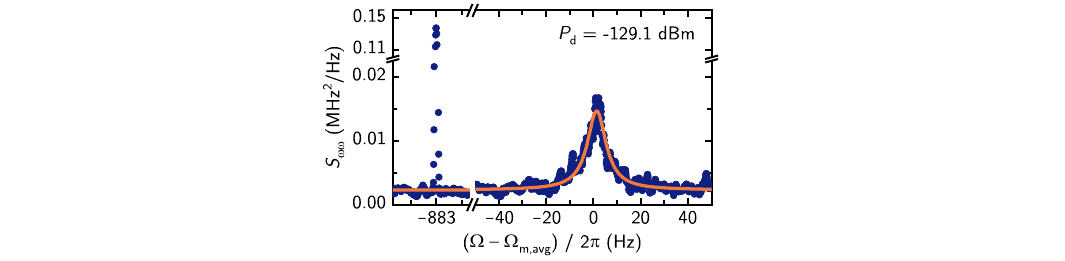}
    \caption{\textbf{Sideband spectroscopy to determine the mechanical parameters.}
    We display a representative spectrum for parameter set III, showing the first mechanical sideband at $\Omega_\mathrm{m,avg}$ and the peak created by the frequency modulation of the drive at roughly $\SI{-883}{\hertz}$. The experimental data are shown as blue dots and fitted with Eq.~\ref{SI:eq:S_ww-fit} (orange line) around the mechanical sideband to determine $\Omega_\mathrm{m}$ and $\Gamma_\mathrm{m}$. By repeating this process for different drive powers, we obtain an average mechanical frequency of $\Omega_\mathrm{m,avg} / 2 \pi = \left( 5.607 483 \pm 0.000 003\right) \si{\mega\hertz}$ and an average linewidth of $\Gamma_\mathrm{m,avg} / 2 \pi = \left( 12 \pm 4\right) \si{\hertz}$.}
    \label{SI:fig:mech-sideband}
\end{figure}

The mechanical frequency $\Omega_\mathrm{m}$, decay rate $\Gamma_\mathrm{m}$, and quality factor $Q_\mathrm{m}$ are determined through sideband spectroscopy. In this technique, a continuous-wave drive is applied to the microwave cavity, and the resulting first-order mechanical sidebands are analyzed. This section provides a comprehensive explanation of this procedure.

Initially, the resonance frequency $\omega_\mathrm{c}$ of the microwave cavity is determined by circle fitting (see Sec.~\ref{SI:sec:cavity-params}). Then, a microwave drive tone with power $P_\mathrm{d}$ is applied on resonance with the cavity at $\omega_\mathrm{d} = \omega_\mathrm{c}$. The drive power is chosen low enough such that the cavity is safely operated in its linear response regime, where we can assume $\mathcal{K} \bar{n}_{\mathrm{c}} \approx 0$ consistent with the photon numbers determined from the Kerr analysis. The transmitted signal is down-converted in a homodyne way, with the local oscillator frequency set equal to the drive frequency ($\omega_\mathrm{LO} = \omega_\mathrm{d}$), and fed into a spectrum analyzer where we observe the appearance of the first mechanical sideband peak at $\Omega_\mathrm{m}$.

Since this measurement can only directly access a voltage spectral density $S_\mathrm{UU}$, we follow the approach introduced in Refs.~\cite{gorodetksy-g0-determination,schmidt-first-paper} and convert into a frequency spectral density $S_\mathrm{\omega \omega}$ by applying a frequency modulation to the drive tone and normalizing the spectral density with respect to its amplitude $S_\mathrm{UU} \left( \Omega_\mathrm{mod} \right)$. Here, $\Omega_\mathrm{mod}$ represents the modulation frequency, which is set to be close but not equal to $\Omega_\mathrm{m}$.

An exemplary resulting spectrum, where the cavity is operated at working point III, is shown in Fig.~\ref{SI:fig:mech-sideband}. In the depicted frequency spectral density, we clearly observe the mechanical sideband as well as a modulation peak. The modulation tone with a frequency deviation of $\Omega_\mathrm{dev} / \left( 2 \pi \right) = \SI{120}{\kilo\hertz}$ is applied at $\Omega_\mathrm{mod} / \left( 2 \pi \right) = \SI{5.606 600}{\mega\hertz}$, which is roughly $\SI{883}{\hertz}$ lower than the mechanical sideband. Assuming a large phonon occupation in the mechanical oscillator $n_\mathrm{m} \approx k_\mathrm{B}T_\mathrm{m} / \hbar \Omega_\mathrm{m} \gg 1$, we can fit the mechanical sideband with \cite{gorodetksy-g0-determination}
\begin{equation}
    S_\mathrm{\omega \omega} \left( \Omega \right) =
    a \frac{\Omega_\mathrm{m} \Gamma_\mathrm{m}}{\left( \Omega^2 - \Omega_\mathrm{m}^2 \right)^2 + \Gamma_\mathrm{m}^2 \Omega^2} + c , \label{SI:eq:S_ww-fit}
\end{equation}
where $a$ is an amplitude, $c$ is a constant offset, $k_\mathrm{B}$ is the Boltzmann constant, and $T_\mathrm{m}$ is the temperature of the mechanical mode. From the fit, we extract the mechanical resonance frequency $\Omega_\mathrm{m}$ and the loss rate $\Gamma_\mathrm{m}$. This process is repeated for various drive powers. In the data set presented here, the drive powers range from $P_\mathrm{d} = \SI{-135.1}{dBm}$ to $\SI{-124.1}{dBm}$. We confirm that the mechanical resonance frequency and loss rate exhibit no clear dependence on the drive power, suggesting that no significant optomechanical sideband heating or cooling of the mechanical mode is induced by the drive. By averaging across all drive powers, we obtain $\Omega_\mathrm{m,avg} / 2 \pi = \left( 5.607 483 \pm 0.000 003\right) \si{\mega\hertz}$ and $\Gamma_\mathrm{m,avg} / 2 \pi = \left( 12 \pm 4\right) \si{\hertz}$ for the given working point corresponding to parameter set III, with the uncertainty given by statistical fluctuations.

Additionally, we estimate the single-photon coupling rate $g_0$ by evaluating the area below the mechanical sideband peak $\braket{\delta\omega_\mathrm{c}^2}$ and assuming thermal equilibrium of the mechanical mode with the cryostat ($T_\mathrm{cryo} = T_\mathrm{m}$) \cite{gorodetksy-g0-determination}. The resulting value for $g_0$ is used to validate the one obtained through electromechanically induced transparency (EMIT) measurements, as described in the upcoming Sec.~\ref{SI:sec:coupling}. For all data sets presented here, both values are in great agreement with another and with previous works (see Refs.~\cite{schmidt-first-paper,luschmann-mech-freq-control}), emphasizing the (self-)consistency of our parameter determination methodology.

\subsection{Single-Photon Coupling Rate} \label{SI:sec:coupling}

\begin{figure}[t]
    \centering
    \includegraphics{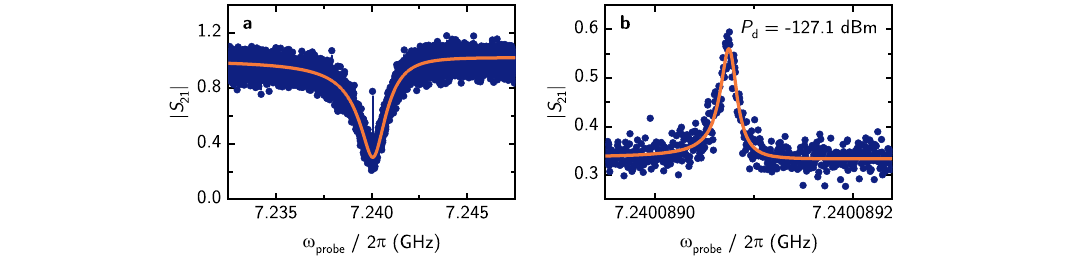}
    \caption{\textbf{Determining $g_0$ by electro-mechanically induced transparency (EMIT).} \textbf{a}, Broad frequency sweep covering the entire cavity resonance. A driving tone of power $P_\mathrm{d} = \SI{-127.1}{dBm}$ is applied on the red sideband at $\omega_\mathrm{d} = \omega_\mathrm{c} - \Omega_\mathrm{m}$. This creates a transparency window in the cavity resonance, showing up here as the single outstanding point close to $\SI{7.240}{\giga\hertz}$. The experimental data are shown as dark blue dots, whereas the orange line is a circle fit.
    \textbf{b}, Narrow high-resolution sweep around the transparency window. With a fine frequency resolution, the Lorentzian lineshape of the EMIT signature is revealed and fitted with Eq.~\ref{SI:eq:s21-emit} to obtain the single-photon coupling rate $g_0$. We repeat this for different drive powers and find an average $g_0 / 2 \pi = \left( 4.69 \pm 0.07\right) \si{\kilo\hertz}$.}
    \label{SI:fig:emit}
\end{figure}

In this section, we explain how the single-photon coupling rate $g_0$ can be determined through electromechanically-induced transparency (EMIT) \cite{Weis-OMIT,Zhou-nanoelectromech,rodrigues-first-nanoelectromech-device,Singh-omit-graphene-steele-natnano2014}. In essence, by applying a sideband drive, a transparency window is created within the cavity resonance, whose amplitude depends on $g_0$. This section explains the details of this process.

Since this measurement extends over a longer period, flux drifts become relevant, which can affect the properties of the microwave resonator, such as its resonance frequency. To mitigate these drifts, we employ an active stabilization technique based on a lock-in amplifier with an integrated proportional-integral-derivative (PID) controller. Technically, we apply a weak continuous-wave stabilization tone to monitor the cavity resonance frequency. The power level of this tone is set sufficiently low to avoid a modification of the cavity response and the mechanical system. The transmitted signal is detected using a lock-in amplifier. The resulting error signal is added to the current used for generating the out-of-plane magnetic field $B_\mathrm{oop}$, which controls $\omega_\mathrm{c}$ and closes the PID loop. This stabilization technique is effective against slow flux drifts occurring on timescales of around $\SI{100}{ms}$ or longer.

In the experiment, a drive tone is first applied at $\omega_\mathrm{d} = \omega_\mathrm{c} - \Omega_\mathrm{m}$ and a broad sweep of the entire cavity resonance is performed with a second weak probe tone. We again choose the drive power low enough to ensure that the cavity is safely in its linear response regime ($\mathcal{K} \bar{n}_{\mathrm{c}} \approx 0$). An exemplary result is illustrated in Fig.~\ref{SI:fig:emit}a for a drive power of $P_\mathrm{d} = \SI{-127.1}{dBm}$ at working point III. The transparency window is only visible as a distinct single-point peak at the center of the cavity resonance. To extract the cavity parameters, this is fitted with a circle fit as described in Sec.~\ref{SI:sec:cavity-params}

Subsequently, we perform a second high-resolution sweep of the cavity scattering response within a narrow range around the EMIT feature to resolve its exact lineshape. The resulting modified Lorentzian lineshape, depicted in Fig.~\ref{SI:fig:emit}b, can be described by \cite{Weis-OMIT,Zhou-nanoelectromech,Singh-omit-graphene-steele-natnano2014}
\begin{equation}
    \left| S_\mathrm{21} \right| = \left| 1 - \frac{\kappa_\mathrm{ext} / 2}{-i \left( \Delta + \Omega_\mathrm{p} \right) + \frac{\kappa}{2} + \frac{g_0^2 \bar{n}_{\mathrm{c}}}{-i \left( \Omega_\mathrm{p} - \Omega_\mathrm{m} \right) + \Gamma_\mathrm{m} / 2}} \right| + c,
    \label{SI:eq:s21-emit}
\end{equation}
where $\Delta = \omega_\mathrm{d} - \omega_\mathrm{c}$ denotes, as in the main text, the detuning of the drive from the cavity resonance, $\Omega_\mathrm{p} = \omega_\mathrm{p} - \omega_\mathrm{d}$ the detuning of the probe from the drive, $c$ a constant background, and $\bar{n}_{\mathrm{c}}$ the intra-cavity photon number. During the fitting procedure, $\bar{n}_{\mathrm{c}}$, $\Gamma_\mathrm{m}$, $\omega_\mathrm{d}$, $\kappa_\mathrm{ext}$, and $\kappa_\mathrm{int}$ remain fixed at previously determined values, while the fit optimizes $c$, $\Omega_\mathrm{m}$, $\omega_\mathrm{c}$, and particularly $g_0$. The (fixed) value of the photon number $\bar{n}_{\mathrm{c}}$ is calculated from the drive power under the assumption that the cavity operates in the linear regime, where \cite{Teufel-strong-coupl-nat2011,bruno-nc-sc-res-applphyslett2015,schmidt-n-ph-calib-applphyslett2018}
\begin{equation}
    \bar{n}_{\mathrm{c}} = \frac{1}{2}\frac{\kappa_\mathrm{ext}}{\Delta^2 + \kappa^2 /4} \frac{P_\mathrm{d}}{\hbar \omega_\mathrm{d}}.
    \label{SI:eq:nc-lin}
\end{equation}
Since an accurate determination of $\bar{n}_{\mathrm{c}}$ requires precise knowledge of $P_\mathrm{d}$, an exact calibration of the input attenuation is necessary. This calibration process is detailed in Sec.~\ref{SI:sec:n-photon-calib}.

We repeat this analysis for various drive powers. In the case of working point III, we vary $P_\mathrm{d}$ from $-134.1$ to $\SI{-109.1}{dBm}$ and obtain a mean value of $g_0 / \left( 2 \pi \right) = \left( 4.69 \pm 0.07 \right) \si{\kilo\hertz}$ for the single-photon coupling, with the uncertainty given by statistical fluctuations. The result is validated by an independent estimate based on sideband spectroscopy, discussed in the previous Sec.~\ref{SI:sec:mech-param}, which underlines the quantitative consistency between the different methods and measurement schemes employed for parameter determination at all examined working points.

\section{Photon Number Calibration}  \label{SI:sec:n-photon-calib}

\begin{figure}[t]
    \centering
    \includegraphics{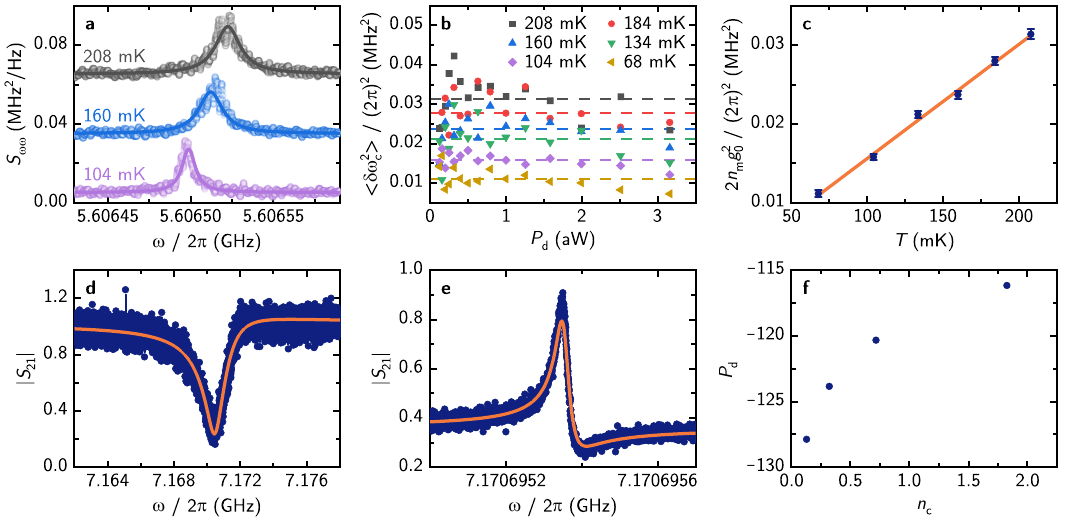}
    \caption{\textbf{Photon number calibration.}
    \textbf{a}, Frequency fluctuations $S_\mathrm{\omega\omega}$ of the cavity showing the first mechanical sideband peak at different temperatures. The power of the red sideband drive is set to $P_\mathrm{d} = \SI{-148.1}{dBm}$. Experimental data are shown in dots, fits with Eq.~\ref{SI:eq:S_ww-fit} as solid lines. The curves are offset by $\SI{0.03}{\mega\hertz^2\per\hertz}$ between each temperature for better visibility.
    \textbf{b}, Areas $\braket{\delta \omega_\mathrm{c}^2}$ below the sideband for varying temperatures $T$ and drive powers $P_\mathrm{d}$. $\braket{\delta \omega_\mathrm{c}^2}$ is determined from the fit results shown in \textbf{a} using Eq.~\ref{SI:eq:d-omg-c-2}. The average value for each temperature is displayed as a dashed line. It shows a clear increase with temperature.
    \textbf{c}, Determination of $g_0$ from the temperature dependence. The average area under the sideband is determined for each temperature and plotted against it. We assume a linear dependence and fit with Eq.~\ref{SI:eq:area-vs-T} to determine a single-photon coupling of $g_0 = \left(4.41 \pm 0.07\right) \si{\kilo\hertz}$ from the slope of the fit.
    \textbf{d}, Broad sweep showing the cavity scattering response in an EMIT measurement where $P_\mathrm{d} = \SI{-120.1}{dBm}$. One can see the red sideband drive as a single outstanding point at roughly $\SI{7.165}{\giga\hertz}$. The experimental data are shown as blue dots, and a circle fit to the data as an orange line.
    \textbf{e}, High-resolution sweep showing the exact EMIT signature. We fit the experimental data (blue dots) with Eq.~\ref{SI:eq:s21-emit} (orange line) to obtain the intracavity photon number $\bar{n}_{\mathrm{c}}$. To be able to do so, we fix the value of $g_0$ to the one determined in \textbf{c}.
    \textbf{f}, Drive power $P_\mathrm{d}$ as a function of the intracavity photon number $\bar{n}_{\mathrm{c}}$ obtained by the fit. We repeatedly fit the EMIT signature for varying driving powers $P_\mathrm{d}$, and calculate the input power of the drive at the sample from the fitted $\bar{n}_{\mathrm{c}}$ using Eq.~\ref{SI:eq:nc-lin}. By comparing to the output power at the microwave source, we find on average a total attenuation of the input line in the dilution refrigerator of $\left(54.4 \pm 0.3\right) \si{\decibel}$.}
    \label{SI:fig:photon-n-calib}
\end{figure}

In order to correctly determine the system parameters and to provide accurate input powers for our numerical simulations and analytical calculations, a precise calibration of the photon number is essential. Conceptually, this is done by first determining $g_0$ at a given working point using a power-independent method and then extracting the photon number in the cavity via EMIT. From this, the power of the drive tone at the sample can be inferred, and by comparing it with the output power of the microwave source, we can determine the total attenuation of the input line and have thus calibrated the photon number. The process is explained in more detail in this section.

The photon number calibration is performed at an in-plane magnetic field of $B_\mathrm{ip} = \SI{10}{\milli\tesla}$. For the determination of the single-photon coupling $g_0$, we follow the temperature-dependent approach introduced among others in Refs.~\cite{gorodetksy-g0-determination,schmidt-first-paper,Regal-mech-motion-cavity-natphys2008,Weber-g0-t-dep-natcomun2016}. Here, the resonance frequency of the cavity is first determined, and based on this, a microwave drive tone is applied at the red sideband at $\omega_\mathrm{d} = \omega_\mathrm{c} - \Omega_\mathrm{m}$, where $\Omega_\mathrm{m} = \SI{5.606451}{\mega\hertz}$. The drive tone is frequency modulated at $\Omega_\mathrm{mod} = \SI{5.605800}{\mega\hertz}$ with a frequency deviation of $\Omega_\mathrm{dev} = \SI{140}{\kilo\hertz}$. The power of the red sideband drive is chosen sufficiently low such that the cavity is operated in its linear response regime ($\mathcal{K} \bar{n}_{\mathrm{c}} \approx 0$) and that optomechanical cooling effects of the mechanical mode can be neglected.

A secondary weak microwave tone at $\omega_\mathrm{stab} = \omega_\mathrm{c} - \SI{0.493549}{\mega\hertz}$ is used to stabilize the cavity against slow flux drifts using an active stabilization technique based on a lock-in amplifier with an integrated PID controller, which is discussed in detail in Sec.~\ref{SI:sec:coupling}. We then heat the sample space to a given elevated temperature and wait for thermal equilibration before performing homodyne sideband spectroscopy. Before and after each measurement, the cavity resonance frequency is determined to check for drifts. This procedure is repeated for varying drive tone power levels and sample space temperatures.

Figure~\ref{SI:fig:photon-n-calib}a shows exemplary sideband spectra at sample space temperatures of $T_\mathrm{cryo} = \SI{208}{\milli\kelvin}$, $\SI{160}{\milli\kelvin}$, and $\SI{104}{\milli\kelvin}$. For all of them, the drive power is set to $P_\mathrm{d} = \SI{-148.1}{dBm}$. As discussed in Sec.~\ref{SI:sec:mech-param}, the original spectrum is converted from a voltage to a frequency spectral density $S_\mathrm{\omega\omega} (\Omega)$. The observed mechanical sideband shape follows Eq.~\ref{SI:eq:S_ww-fit}, which we fit to the experimental spectra to determine the area under the curve $\braket{\delta \omega_\mathrm{c}^2}$. This is linked to the phonon occupation $n_\mathrm{m}$ of the mechanical mode via the single-photon coupling rate \cite{gorodetksy-g0-determination}
\begin{equation}
    \braket{\delta \omega_\mathrm{c}^2} = \int_{-\infty}^{+\infty} S_\mathrm{\omega\omega} \left( \Omega \right) \mathrm{\frac{d \Omega}{2 \pi}} = S_\mathrm{\omega\omega} \left( \Omega_\mathrm{m} \right) \frac{\Gamma_\mathrm{m}}{2} = 2 n_\mathrm{m} g_0^2 . \label{SI:eq:d-omg-c-2}
\end{equation}
Figure~\ref{SI:fig:photon-n-calib}b shows the dependence of the mechanical sideband area $\braket{\delta \omega_\mathrm{c}^2}$ on the red sideband drive power for different temperatures. It slightly varies around a mean value, confirming that the drive power is sufficiently low. Only at very high powers do slight optomechanical cooling effects begin to appear due to a minimal detuning from cavity resonance, which, however, only have an insignificant effect on the measurement results.

Since the mechanical sidebands are recorded at different temperatures, there is no need to assume that the mechanical mode thermalizes with the sample stage in the dilution refrigerator. Instead, $g_0$ can be determined from the slope of the linear temperature dependence of the area below the sideband (averaged over all drive powers) \cite{schmidt-first-paper,Regal-mech-motion-cavity-natphys2008}
\begin{equation}
    2 n_\mathrm{m} g_0^2 = s_\mathrm{T} T + c_\mathrm{ba} , \label{SI:eq:area-vs-T}
\end{equation}
where $c_\mathrm{ba}$ is a bath temperature and $s_\mathrm{T} = (2 k_\mathrm{B} g_0^2) / (\hbar \Omega_\mathrm{m})$ represents the slope of the linear dependence with $k_\mathrm{B}$ the Boltzmann constant. From the fit displayed in Fig.~\ref{SI:fig:photon-n-calib}c we obtain a single-photon coupling rate of $g_0 = \left(4.41 \pm 0.07\right) \si{\kilo\hertz}$.

Based on this, we can now determine the photon number $\bar{n}_{\mathrm{c}}$ inside the cavity using EMIT measurements. Here, the cavity is also first stabilized against flux drifts using a lock-in amplifier with an integrated PID controller as discussed previously. A drive tone is then applied at the red sideband $\omega_\mathrm{d} = \omega_\mathrm{c} - \Omega_\mathrm{m}$. With a second weak probe tone, the scattering response of the cavity is measured and the cavity parameters are extracted using a circle fit (see Fig.~\ref{SI:fig:photon-n-calib}d for $P_\mathrm{d} = \SI{-120.1}{dBm}$). A narrow zoom into the center of the cavity resonance reveals the transparency window, which is fitted using Eq.~\ref{SI:eq:s21-emit}, as shown in Fig.~\ref{SI:fig:photon-n-calib}e. Unlike for the determination of the coupling rate in Sec.~\ref{SI:sec:coupling}, this time we fix $\Gamma_\mathrm{m}$, $\omega_\mathrm{d}$, $\kappa_\mathrm{ext}$, $\kappa_\mathrm{int}$, and $g_0$ to the previously determined values for the fitting, while optimizing $c$, $\Omega_\mathrm{m}$, $\omega_\mathrm{c}$, and $\bar{n}_{\mathrm{c}}$. Here, $g_0$ is fixed to the value obtained through temperature-dependent sideband spectroscopy. From the value obtained for the intra-cavity photon number $\bar{n}_{\mathrm{c}}$, we can calculate the drive power at the sample by solving Eq.~\ref{SI:eq:nc-lin} for $P_\mathrm{d}$ and assuming that the cavity is operated in the linear regime ($\mathcal{K} \bar{n}_{\mathrm{c}} \approx 0$)
\begin{equation}
    P_\mathrm{d} = 2 \bar{n}_{\mathrm{c}} \frac{\Delta^2 + \kappa^2/4}{\kappa_\mathrm{ext}} \hbar \omega_\mathrm{d} .
\end{equation}
This is repeated for various drive powers. Figure~\ref{SI:fig:photon-n-calib}f shows the values obtained. By comparing the output power of the microwave source with the drive power at the sample, we determine on average an input line attenuation of $\left(54.4 \pm 0.3\right) \si{\decibel}$, thus completing the power calibration.

\section{Pulsed Measurement Technique} \label{SI:sec:pulsed-transient}

\begin{figure}[t]
    \centering
    \includegraphics{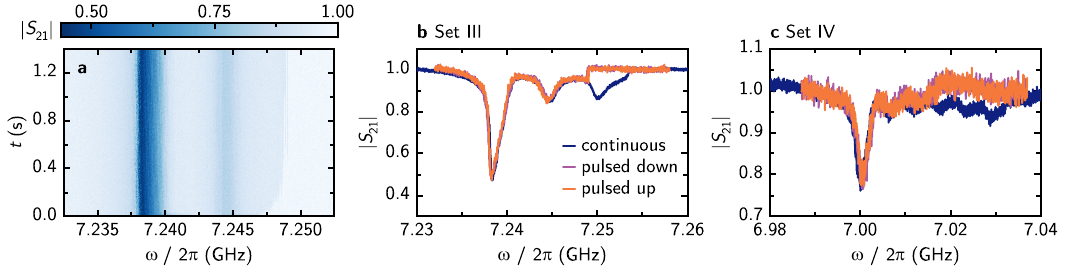}
    \caption{\textbf{Pulsed Measurement Technique.}
    \textbf{a}, Scattering response as a function of the probe frequency $\omega$ and time $t$ obtained using the pulsed measurement technique. The data shown corresponds to parameter set III at an input power of $P_\mathrm{d} = \SI{-114.6}{dBm}$. The frequencies are measured in ascending order.
    \textbf{b}, Comparison of different sweep directions and measurement techniques for parameter set III and $P_\mathrm{d} = \SI{-114.6}{dBm}$. From the data shown in \textbf{a}, we extract the frequency slice at the latest measured time (labeled as \textit{pulsed up}). We repeat this for a pulse sequence that starts at high and goes to low frequencies (labeled as \textit{pulsed down}). The consistency of the observed response across sweep directions demonstrates that our pulsed measurement technique ensures that the system is always initialized in thermal equilibrium before each frequency measurement. In contrast, a conventional continuous sweep with no waiting times between frequencies and no time-resolved information reveals an additional absorption dip on the blue sideband, which we attribute to transient effects.
    \textbf{c}, Comparison between sweep directions and measurement techniques for parameter set IV and $P_\mathrm{d} = \SI{-120.6}{dBm}$, showing similar trends. Due to the large $g_0$ and $\mathcal{K}$ values, we observe five side absorption dips with the conventional continuous sweep and only two with our pulsed scheme.}
    \label{SI:fig:pulsed-S21-time-freq}
\end{figure}

As discussed in the main text and illustrated there in Fig.~\ref{main:fig3:s21comp}a,b, we employ a pulsed measurement scheme to ensure that the system reaches its steady state for each examined detuning, while also allowing equilibration with its thermal environment between pulses. This ensures that we do not need to account for transient dynamics originating from the starting conditions of the mechanical state. Effects related to transient dynamics are not included in our theoretical model. In this section, we show the experimentally observed difference between the situation of deliberately prepared initial conditions and signatures that can be associated with the transient dynamics.

Our pulsed approach introduces an additional dimension to the retrieved information, namely the time evolution $t$, which is not accessible in conventional vector network analyzer (VNA) measurements. A representative measurement at a given input power of $P_\mathrm{d} = \SI{-114.6}{dBm}$, corresponding to $n_\mathrm{in} / n_{\mathrm{in,crit}} = 2.77$, is shown in Fig.~\ref{SI:fig:pulsed-S21-time-freq}a for parameter set III. The same data is also presented in Fig.~\ref{main:fig4:colormap-slices}III.c in the main text. Here, one can see that the primary cavity resonance remains approximately constant over time, as does the single side absorption dip visible slightly below $\SI{7.245}{\giga\hertz}$. However, a distinct time evolution is observed at the high-frequency end of the cavity response, where up to roughly $\SI{0.8}{\second}$ are required for the dip to fully emerge, indicating that the system transitioned into its steady state. This observation underlines the necessity of the pulsed measurement technique to accurately capture the steady state response in the experiment.

From this dataset, we extract the frequency slice at the latest measured time ($t = \SI{1.4}{\second}$). The resulting spectrum is shown in Fig.~\ref{SI:fig:pulsed-S21-time-freq}b as the orange line for a sweep, where the frequency of the pulses starts at the lowest and ends at the highest measured frequency. A complementary pulsed sweep in the opposite direction, from the highest to the lowest frequency, is indicated by the purple line. Both measurements show an identical frequency dependence. For comparison, we perform a continuous sweep using a standard vector network analyzer, where each frequency tone is applied and averaged for $\SI{10}{\milli\second}$ without any waiting time in between, and the frequency is swept from low to high. The result of this conventional method is displayed in Fig.~\ref{SI:fig:pulsed-S21-time-freq}b as a dark blue line. The same is displayed in Fig.~\ref{SI:fig:pulsed-S21-time-freq}c for a different working point, namely parameter set IV, and $P_\mathrm{d} = \SI{-120.6}{dBm}$ ($n_\mathrm{in} / n_{\mathrm{in,crit}} = 2.88$). This corresponds to the highest applied power for the largest value of $g_0$ and $\mathcal{K}$ examined in this work. For Fig.~\ref{SI:fig:pulsed-S21-time-freq}b and c we observe that, with the pulsed scheme, the sweep direction does not affect the outcome, confirming that each frequency point is indeed completely independent of the others. Consequently, the measurement could be performed in any arbitrary frequency order without affecting the result.

However, a significant discrepancy arises when comparing the pulsed scheme to the conventional continuous sweep. The latter reveals additional absorption dips extending much further into the blue sideband than predicted by out model. For the parameters in Fig.~\ref{SI:fig:pulsed-S21-time-freq}b, this manifests as a single additional dip. But at higher powers and larger values of $g_0$ and $\mathcal{K}$ as in Fig.~\ref{SI:fig:pulsed-S21-time-freq}c, the number of dips more than doubles in the continuous sweep compared to the pulsed method, producing a significantly different picture. We attribute these additional absorption dips to transient behavior in which the self-sustained oscillations of the mechanical subsystem do not fully ring down between the detunings and thus can be carried on to larger detunings, given that the frequency is swept in ascending order. A frequency sweep in descending order does not observe these additional dips.

These transient features are beyond the scope of our model, which essentially computes the steady state response. This confirms that the pulsed measurement scheme indeed ensures that only the steady state is probed and transient effects are eliminated. Although the transient dynamics of the system are of interest in themselves, we focus on the intrinsic dynamics in this work.

\section{Additional Parameter Sets} \label{SI:sec:add-param-set}

\begin{table*}[b]
\caption{\textbf{Experimental system parameters for all working points at which measurements were taken.}
All parameters are determined independently and used as input for numerical simulations of the scattering response. Note that we can cover values in $g_0$ and $\mathcal{K}$ over a range of two orders of magnitude. We attribute irregular variations of the mechanical loss rate $\Gamma_\mathrm{m}$ between the parameter sets to (i) the general uncertainty in the determination of $\Gamma_\mathrm{m}$ and (ii), especially for set IV, the increase in flux noise susceptibility.}
\label{SI:tab:param-sets-all} 
\begin{ruledtabular}
\begin{tabular}{lllll}
 & Set I & Set II & Set III & Set IV \\ \hline
$\omega_\mathrm{c}$ (GHz) & $7.330$ & $7.310$ & $7.241$ & $7.006$ \\
$\partial\omega_\mathrm{c} / \partial\Phi_\mathrm{ext} \, (\si{\giga\hertz} / \Phi_0)$ & $0.15$ & $0.31$ & $0.83$ & $3.22$ \\
$\kappa_\mathrm{int}$ (MHz) & $0.57$ & $0.60$ & $0.68$ & $2.33$ \\
$\kappa_\mathrm{ext}$ (MHz) & $2.00$ & $1.72$ & $1.64$ & $1.55$ \\
$\mathcal{K}$ (kHz) & $16$ & $20$ & $70$ & $1.4 \times 10^3$ \\
$\bar{n}_\mathrm{c,crit}$ & $92$ & $ 67$ & $19$ & $1.6$ \\
$\Omega_\mathrm{m}$ (MHz) & $5.607 716$ & $5.607 653$ & $5.607 483$ & $5.607 110$ \\
$\Gamma_\mathrm{m}$ (Hz) & $11$ & $14$ & $12$ & $6$ \\
$g_0$ (kHz) & $0.76$ & $1.95$ & $4.69$ & $18.4$ \\
$g_0 / \mathcal{K}$ & $0.05$ & $0.10$ & $0.07$ & $0.01$ \\
\end{tabular}
\end{ruledtabular}
\end{table*}

\begin{figure}[t]
    \centering
    \includegraphics{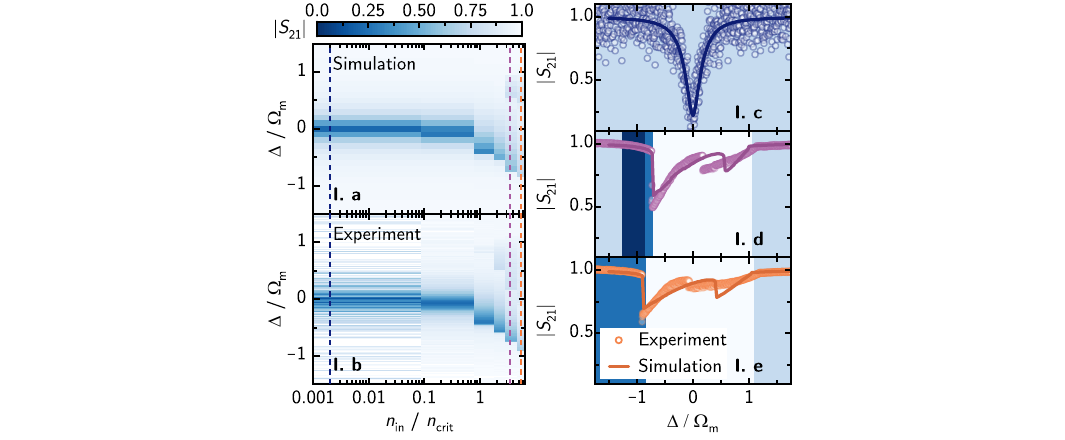}
    \caption{\textbf{Cavity scattering response for parameter set I in table \ref{SI:tab:param-sets-all}.}
    The two color plots compare simulation \textbf{I.a} and experiment \textbf{I.b} for a wide set of probe powers. The line cuts in \textbf{I.c,d,e} show the experimental data as points and numerical simulation as lines for a single probe power increasing from $n_\mathrm{in} / n_{\mathrm{in,crit}} = 0.002$ \textbf{I.c}, over $3.54$ in \textbf{I.d} to $5.61$ in \textbf{I.e}, which correspond to $P_\mathrm{d} = \SI{-139.6}{dBm}$, $\SI{-106.6}{dBm}$, and $\SI{-104.6}{dBm}$. These powers are indicated in \textbf{a,b} by the dashed vertical lines. The background is colored according to the stability regions.}
    \label{SI:fig:colormap-slices-set-I-IV}
\end{figure}

We performed all measurements for a set of four different working points, as indicated by the orange stars in Fig.~\ref{SI:fig:coil-kerr-sweep}a. For each working point, the system parameters were determined using the methods described previously. A summary of these parameters is provided in table~\ref{SI:tab:param-sets-all}, demonstrating that our system is capable of covering a wide range of single-photon coupling rates and Kerr nonlinearities $\mathcal{K}$, spanning nearly two orders of magnitude, i.e.~from $g_0 / (2\pi) = \SI{0.76}{\kilo\hertz}$ to $\SI{18.4}{\kilo\hertz}$. Based on these parameters, we simulate the scattering response of the cavity. While the main text focuses on parameter sets II, III, and IV, this section presents further experiments for parameter set I, which are not presented in the main text.

Fig.~\ref{SI:fig:colormap-slices-set-I-IV}a(b) shows the simulation (experiment) for parameter set I for a wide range of input powers as a heatmap. Single power slices are displayed in Fig.~\ref{SI:fig:colormap-slices-set-I-IV}c,d,e. For this dataset, we observe slight quantitative discrepancies between the simulation and experimental data, in particular for the peak location of the mechanical signature. However, the boundary between the stability regions \textbf{i} and \textbf{ii} is correctly predicted. We attribute this discrepancy to the following factors. The determined value of $g_0$ has the highest uncertainty in this dataset, which originates from the relatively low magnitude of $g_0$. As the observation of the nonlinear signatures critically depends on this value, small deviations and uncertainties in the independently determined quantity $g_0$ result in discrepancies between model and experiment. In addition, the relatively high drive powers can induce the occupancy of thermal microwave photons in the microwave resonator, which is not accounted for in our model.

However, the qualitative agreement between experiment and simulation still remains exceptional. In comparison to set IV, the values of both $g_0$ and $\mathcal{K}$ for set I are almost two orders of magnitude lower. Despite these vastly different values, the simulations still describe the experimental data to great agreement.

\section{Comparison to Related Works} \label{SI:sec:compare-other-works}

\begin{figure}[t]
    \centering
    \includegraphics{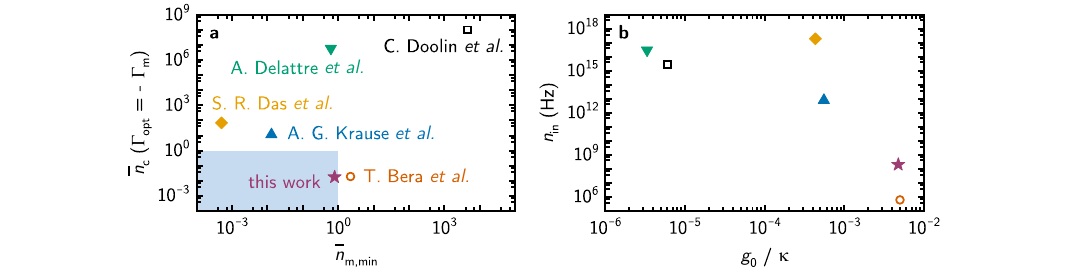}
    \caption{\textbf{Comparison to previous works.}
    We benchmark our device against other earlier experiments reported in literature in terms of single-photon coupling rate $g_0$, loss rate $\kappa$ of the cavity, input photon flux of the drive $n_\mathrm{in}$ and mechanical resonance frequency $\Omega_\mathrm{m}$. We compare to the works from C. Doolin \textit{et al.} \cite{Doolin-nonlin-opto-stat-pra2014}, A. G. Krause \textit{et al.} \cite{krause2015}, S. R. Das \textit{et al.} \cite{srdas2023}, T. Bera \textit{et al.} \cite{Bera-instab-natcomun2024} and A. Delattre \textit{et al.} \cite{Delattre-self-sust-destr-prres2024}. The data points from the other studies are calculated using the parameter values specified in the respective study. For this work, we used parameter set IV. Hollow symbols indicate sideband-unresolved systems, while the filled symbols correspond to sideband-resolved systems.
    \textbf{a}, We compare the intra-cavity photon number $\bar{n}_\mathrm{c}(\Gamma_\mathrm{opt} = -\Gamma_\mathrm{m})$ required for mechanical instability and the minimal phonon number $\bar{n}_\mathrm{m,min}$ achievable by red-sideband cooling. The light-blue region indicates the desired regime to realize a quantum nonlinear mechanical system. It is defined by the conditions $\bar{n}_\mathrm{c}(\Gamma_\mathrm{opt} = -\Gamma_\mathrm{m}) \leq 1$ and $\bar{n}_\mathrm{m,min} \leq 1$.
    \textbf{b}, Comparison of the single-photon coupling over the loss rate $g_0 / \kappa$ and the input photon flux $n_\mathbf{in}$ required to observe mechanical instabilities. 
    In both benchmarks, our system outperform previous works. It reaches a regime where one can prepare a mechanical system in a non-classical state and subsequently drive it into a instable region without destroying the quantum state.}    
    \label{SI:fig:lit-compare}
\end{figure}

A future goal, to which our work is paving the way, is the exploration of quantum physics with a macroscopic mechanical system such as our nanostring, which operated deep in its nonlinear regime. Here, a potential experiment is to initially prepare the mechanical resonator in a non-classical quantum state and subsequently probe its evolution while being driven into its unstable, nonlinear regime. To be able to realize such or similar experiments, two fundamental requirements must be met: (i) the ability to prepare the mechanical system in or near its ground state, and (ii) the possibility to drive the system into instability at very low driving powers or cavity photon numbers, such that the initial quantum state is not destroyed by the drive at the onset of nonlinear dynamics. This section benchmarks our optomechanical system against previous works, in which mechanical instabilities have been reported, in terms of the possibility of performing experiments as described before in the future. This is done by identifying the photon number threshold for instability of the mechanical system and the minimal phonon occupation achievable under sideband cooling.

We assume that the mechanical system transitions into the unstable regime when the optomechanical damping rate $\Gamma_\mathrm{opt}$ creates an anti-damping that effectively cancels out the intrinsic damping rate $\Gamma_\mathrm{m}$ of the mechanics ($\Gamma_\mathrm{opt} = - \Gamma_\mathrm{m}$). For simplicity, $\Gamma_\mathrm{opt}$ is calculated for the case of a blue sideband drive $\Delta = \Omega_\mathrm{m}$ on a linear cavity as the threshold is only marginally influenced by the Kerr nonlinearity. It is given by\cite{aspelmeyer-cavity-optomechanics}
\begin{equation}
    \bar{n}_\mathrm{c} \left( \Gamma_\mathrm{opt} = - \Gamma_\mathrm{m} \right) =
    \frac{1 + \bar{n}_\mathrm{m,lin}}{C_0} ,
\end{equation}
where the single-photon cooperativity is $C_0 = \tfrac{4 g_0^2}{\Gamma_\mathrm{m} \kappa}$ and the minimum phonon number of a linearized system in the absence of a thermal bath is $\bar{n}_\mathrm{m,lin} = (\tfrac{\kappa}{4 \Omega_\mathrm{m}})^2$. Ideally, this threshold should be reached at cavity occupations of order unity or below.

To prepare the mechanical oscillator initially in a non-classical state, it needs to be cooled into or near its quantum ground state. To quantify this, we calculate the minimum phonon occupation of a linear optomechanical system that can be achieved using red-sideband cooling ($\Delta = -\Omega_\mathrm{m}$). The case of a Kerr nonlinear cavity again modifies the result only slightly and is therefore neglected. We assume the presence of a thermal bath at a given temperature $T$, which leads to a thermal occupation $\bar{n}_\mathrm{th}$ of the mechanical system. The cooling power is assumed to be limited by the cavity occupation, at which it bifurcates, thus defining the lower bound on the achievable phonon number as\cite{aspelmeyer-cavity-optomechanics,nico-nonlinear-cooling}
\begin{equation}
    \bar{n}_\mathrm{m,min} = \frac{4 g_0^2 \kappa^2 + \sqrt{3} \Gamma_\mathrm{m} \mathcal{K}_\mathrm{eff} \bar{n}_\mathrm{th} \left(16 \Omega_\mathrm{m}^2 + \kappa^2 \right)}{64 g_0^2 \Omega_\mathrm{m}^2 + \sqrt{3} \Gamma_\mathrm{m} \mathcal{K}_\mathrm{eff} \left(16 \Omega_\mathrm{m}^2 + \kappa^2 \right)} ,
\end{equation}
where $\mathcal{K}_\mathrm{eff} = \mathcal{K} + \frac{2 g_0^2 \Omega_\mathrm{m}}{\Omega_\mathrm{m}^2 + \Gamma_\mathrm{m}^2 / 4}$ is the effective Kerr given by the sum of the cavity and the mechanical Kerr. This is especially relevant for systems with a finite Kerr nonlinearity of the cavity, but also limits linear cavity systems via the mechanical Kerr.

To provide a fair comparison with previous works, we plot the instability threshold photon number against the minimal reachable phonon number in Fig.~\ref{SI:fig:lit-compare}, assuming a common temperature of $\SI{20}{\milli\kelvin}$. We find that while all previous experiments remain outside the region, where a quantum nonlinear mechanical system can potentially be realized, our device is the first to reach it. This should enable experiments in this directions in the near future.

\section{Theoretical Model} \label{si:sec:theory-model}

The interaction of an intrinsically nonlinear cavity coupled to a mechanical oscillator is given by the following Hamiltonian \cite{nation2008,nico-nonlinear-cooling},
\begin{equation}\label{si:ham1}
    \hat{H}_{\mathrm{tot}} = \omega_\mathrm{c}
    \hat{a}^{\dag} \hat{a} + \Omega_\mathrm{m} \hat{b}^{\dag} \hat{b} -\frac{\mathcal{K}}{2} \hat{a}^{\dag} \hat{a}^{\dag} \hat{a} \hat{a} +  g_0 \hat{a}^{\dag}\hat{a} (\hat{b} + \hat{b}^{\dag}) + \hat{H}_{\mathrm{drive}} ,
\end{equation}
where $\omega_\mathrm{c}$, $\Omega_\mathrm{m}$ are the resonance frequencies of the optical and mechanical modes. The annihilation and creation operators for the optical (mechanical mode) are given by $\hat{a}(\hat{b})$ and $\hat{a}^{\dag}(\hat{b}^{\dag})$, respectively. $\mathcal{K}>0$ is the Kerr nonlinearity of the cavity mode and $g_0 = g x_{\mathrm{zpf}}$ denotes the bare optomechanical coupling strength, where $x_{\mathrm{zpf}}$ is the zero point motion of the mechanical resonator mode and $g = \partial \omega_c/\partial x$ is the optical frequency shift per displacement. $\hat{H}_{\mathrm{drive}}$ is the Hamiltonian of the input drive and is given by $\hat{H}_{\mathrm{drive}} = -i \sqrt{\kappa_{\mathrm{ext}}/2}(\alpha_\mathrm{p} e^{-i\omega_\mathrm{p} t} \hat{a}^{\dag} - h.c.)$ with $\omega_\mathrm{p}$ being the driving frequency of the pump, $\alpha_\mathrm{p}$ being the pump amplitude, and $\kappa_{\mathrm{ext}}$ being the decay rate associated with the external losses in the cavity. The factor of $1/2$ with $\kappa_{\mathrm{ext}}$ can be attributed to the bidirectional coupling of the transmission line to the microwave cavity \cite{Wang_2021-hangermode}.

In the frame rotating with respect to the input drive's frequency $\omega_\mathrm{p}$, we obtain the following Hamiltonian \cite{nico-nonlinear-cooling}, 
\begin{equation}\label{si:ham2}
    \hat{H} = -\Delta \hat{a}^{\dag} \hat{a} + \Omega_\mathrm{m} \hat{b}^{\dag} \hat{b} -\frac{\mathcal{K}}{2} \hat{a}^{\dag} \hat{a}^{\dag} \hat{a} \hat{a} + g_0 \hat{a}^{\dag} \hat{a}(\hat{b} + \hat{b}^{\dag}),
\end{equation}
with the bare detuning $\Delta = \omega_\mathrm{p} - \omega_c$.
In the following sections, given the Hamiltonian in Eq.~\ref{si:ham2}, we will calculate the complex scattering response of the cavity in the stable and unstable regimes. We will examine both analytical and numerical methods with the goal to obtain a scattering response which accurately describes a microwave transmission experiment.

\subsection{Analytics}\label{SI: analytial model}

Using the Hamiltonian given in Eq.~\ref{si:ham2}, we obtain the following equations of motion for the cavity and mechanical mode using standard input-output theory \cite{gardiner1985}, 
\begin{align}
    \mathrm{\frac{d}{dt}} \hat{a} &= i \Delta \hat{a} -\frac{\kappa}{2} \hat{a} + i\mathcal{K} \hat{a}^{\dag} \hat{a} \hat{a} - ig_0 \hat{a}(\hat{b}+\hat{b}^{\dag}) - \sqrt{\frac{\kappa_\mathrm{ext}}{2}} \alpha_\mathrm{p}-\sqrt{\kappa} \hat{a}_\mathrm{in} ,\\
    \mathrm{\frac{d}{dt}} \hat{b} &= -(i\Omega_\mathrm{m} +\frac{\Gamma_\mathrm{m}}{2}) \hat{b} - ig_0 \hat{a}^{\dag} \hat{a} - \sqrt{\Gamma_{\mathrm{m}}} \hat{b}_{\mathrm{in}},
\end{align}
where $\kappa$ and $\Gamma_{\mathrm{m}}$ denote the dissipation rates associated with the coupling of the cavity and mechanical mode to their respective baths. Here, $\kappa = \kappa_{\mathrm{ext}}+\kappa_{\mathrm{int}}$ is the total decay rate associated with the external and internal losses of the cavity, and $\hat{a}_\mathrm{in} ( \hat{b}_\mathrm{in})$ are the input noise operators associated with the cavity (mechanical) mode. The cavity noise operator is defined as  $\hat{a}_\mathrm{in} = \sqrt{\frac{\kappa_\mathrm{ext}}{2\kappa}}(\hat{\xi}_\mathrm{L,in}+ \hat{\xi}_\mathrm{R,in}) + \sqrt{\frac{\kappa_\mathrm{int}}{\kappa}} \hat{\xi}_{\mathrm{int}}$ where $\hat{\xi}_\mathrm{L,in}, \hat{\xi}_\mathrm{R,in}$ are the noise channels associated with the coupling of the device to the transmission line and $\hat{\xi}_{\mathrm{int}}$ denotes the noise channel associated with internal losses. $\alpha_\mathrm{p}$ is the coherent drive injected through the external port. It is important to note that the transmission line is attached to the resonator in a hanger-type configuration where the input drive is applied only in one direction. This leads to a factor of 1/2 with the decay rate $\kappa_{\mathrm{ext}}$. In contrast, the prefactor $1/2$ can be removed if the device is operated in a reflection geometry \cite{Wang_2021-hangermode}. Such a convention has also been used in other works on optomechanics, e.g., Ref.\,\cite{krause2015}.

Under the assumption of a strong pump power, we separate our system dynamics into an average and a fluctuation part, $\hat{a}=\alpha+\delta\hat{a}$ and $\hat{b} = \beta + \delta\hat{b}$, for both the cavity and the mechanics, respectively, to obtain the following nonlinear equations of motion
   \begin{align}\label{cavity-mode-eq}
    \mathrm{\frac{d}{dt}} \alpha &= (i\Delta - \frac{\kappa}{2}) \alpha - ig_0(\beta+\beta^*) \alpha + i\mathcal{K}|\alpha|^2  \alpha - \sqrt{\frac{\kappa_{\mathrm{ext}}}{2}} \alpha_\mathrm{p},  \\ \label{mechanical-mode-eq}
    \mathrm{\frac{d}{dt}} \beta &= -\left(i\Omega_\mathrm{m}+\frac{\Gamma_\mathrm{m}}{2}\right) \beta - ig_0 |\alpha|^2,
\end{align}
where the classical amplitudes $\alpha$ and $\beta$ describe the system dynamics. Here $\alpha_\mathrm{p}$ is the coherent drive injected through the external port. The above coupled set of equations can be solved in the long-time limit, assuming that $|\alpha|^2$ is the steady state photon occupation number. When the mechanical subsystem is stable (that means the total damping is positive), the steady state response of the mechanical mode is obtained by solving Eq.~\ref{mechanical-mode-eq} as to find
\begin{equation}
    \beta = - \frac{i g_0 |\alpha|^2}{i\Omega_\mathrm{m}+\Gamma_\mathrm{m}/2} ,
\end{equation}
which we insert into Eq.~\ref{cavity-mode-eq} to obtain the following equation for the classical amplitude of the cavity mode
\begin{equation} \label{cavity-mode-eq2}
    \mathrm{\frac{d}{dt}} \alpha = (i\Delta - \frac{\kappa}{2}) \alpha + i\mathcal{K}_{\mathrm{eff}}|\alpha|^2  \alpha - \sqrt{\frac{\kappa_{\mathrm{ext}}}{2}} \alpha_\mathrm{p} .
\end{equation}
Here, the effective Kerr constant,
\begin{equation}
    \mathcal{K}_{\mathrm{eff}} = \mathcal{K} + \frac{2 g_0^2 \Omega_\mathrm{m}}{\Omega_\mathrm{m}^2+{\Gamma^2_\mathrm{m}}/4} ,
\end{equation}
is composed of the intrinsic cavity nonlinearity $\mathcal{K}$ and the nonlinearity induced due to the optomechanical interaction \cite{aldana2013}, also referred to as the mechanical Kerr $\mathcal{K}_{\mathrm{m}}$ in this work. The intracavity photon occupation number $\bar{n}_{\mathrm{c}}$ can be obtained by multiplying the steady state solution of Eq.~\ref{cavity-mode-eq2} with its complex conjugate
\begin{equation}\label{photon-no-eq}
\bar{n}_{\mathrm{c}} = \frac{\kappa_{\mathrm{ext}}}{2} n_{\mathrm{in}}  \frac{1}{(\Delta+\mathcal{K}_{\mathrm{eff}}\bar{n}_{\mathrm{c}})^2+\kappa^2/4} ,
\end{equation}
where $n_{\mathrm{in}} = |\alpha_\mathrm{p}|^2$ is the input power (in units of photons per second). As expected from a cubic equation, we obtain one or three real solutions for the photon number, depending on the input drive power $n_{\mathrm{in}}$. For low input driving powers, a single real solution exists. However, as the input power is increased, the system undergoes a bifurcation and enters a bistable regime, where the photon occupation has three possible solutions. The bifurcation occurs when the first and second derivative of Eq.~\ref{photon-no-eq} with respect to $\bar{n}_{\mathrm{c}}$ vanishes. This allows us to solve for the critical value of detuning at $\Delta_\mathrm{crit} = -\sqrt{3}\kappa/2$ and the photon number $\bar{n}_\mathrm{c,crit} = \kappa/\sqrt{3} \mathcal{K}_{\mathrm{eff}}$. Further, solving for the input power at this critical parameters gives us the critical input power $n_{\mathrm{in,crit}}$ for our setup \cite{laflamme2011,nico-nonlinear-cooling}, which is given by
\begin{equation}
    n_{\mathrm{in,crit}} = \frac{2 \kappa^3}{3\sqrt{3} \kappa_{\mathrm{ext}} \mathcal{K}_{\mathrm{eff}}}.
\end{equation}
As can be seen from the above equation, the critical input power $n_{\mathrm{in,crit}}$ only depends on the decay rates of the cavity and the effective Kerr constant $\mathcal{K}_{\mathrm{eff}}=\mathcal{K}+\mathcal{K}_{\mathrm{m}}$. 
Under the assumption that $|\alpha|^2$ can be approximated as the solution of Eq.~\ref{photon-no-eq} for the photon occupation $\bar{n}_{\mathrm{c}}$, the solution for the cavity amplitude $\alpha$ can be obtained by Fourier transforming Eq.~\ref{cavity-mode-eq2}
\begin{equation}\label{app:s21-eq}
    \alpha[\omega]= \frac{-\sqrt{\kappa_{\mathrm{ext}}/2}\alpha_\mathrm{p}}{-i(\omega+\Delta +\mathcal{K}_{\mathrm{eff}} \bar{n}_{\mathrm{c}})+\kappa/2}.
\end{equation}
In the experiment, a single-tone microwave transmission measurement is performed. In other words, there is only a single strong tone, which is used both as the pump and the probe tone. Obtaining the response as in Eq.~\ref{app:s21-eq} models a two-tone measurement, where a strong pump is applied at a chosen frequency, and a weak probe of frequency $\omega$ scans over the parameter space. To obtain the scattering response measured in the experiment with only a single tone, we remove the second weak probe tone, denoted by $\omega$, from Eq.~\ref{app:s21-eq} by setting its frequency to zero. Employing the input-output relation gives us the scattering response of the cavity mode
\begin{equation} \label{si:scatresponse1}
    S_{21}[\omega=0] = 1- \frac{\kappa_{\mathrm{ext}}}{2} \frac{1}{-i(\Delta+\mathcal{K}_{\mathrm{eff}} \bar{n}_{\mathrm{c}})+\kappa/2} .
\end{equation}
The scattering response in Eq.~\ref{si:scatresponse1} describes the cavity response in the regime where the mechanical mode is assumed to be oscillating with a small amplitude. However, backaction effects can lead to heating of the mechanical mode up to the point of dynamical instability, where it starts to oscillate with a high steady state amplitude. To solve for the cavity response in this unstable regime, we start by taking the usual ansatz for the mechanical mode \cite{marquardt2006}
\begin{equation}
    \beta = \bar{\beta}+\frac{B}{x_{\mathrm{zpf}}}e^{-i\Omega_\mathrm{m} t}e^{-i\phi} ,
    \label{si:eq:ansatz-mechannics-beta}
\end{equation}
where the first term $\bar{\beta}$ denotes the static displacement, and the second term denotes the self-oscillatory motion of the mechanical mode. In the second term, the mechanical mode is assumed to oscillate at its resonance frequency $\Omega_\mathrm{m}$ with a relative phase of $\phi$ and an oscillation amplitude $B$. Plugging in this ansatz into Eqs.~\ref{cavity-mode-eq} and \ref{mechanical-mode-eq} gives the following solution for the cavity amplitude $\alpha$ in the time domain
\begin{equation}\label{cavity-sol-time}
    \alpha(t) = - \sqrt{\frac{\kappa_{\mathrm{ext}}}{2}} \alpha_\mathrm{p} e^{-iz_1 \sin(\Omega_\mathrm{m} t+\phi)} \sum_{n} \frac{J_n(z_1) e^{in(\Omega_\mathrm{m} t +\phi)}}{-i(\Delta+\mathcal{K}_{\mathrm{eff}}\bar{n}_{\mathrm{c}}-n\Omega_{\mathrm{m}})+\kappa/2} .
\end{equation}
Here, $z_1=2Bg/\Omega_\mathrm{m}$ is the argument of the Bessel function of the first kind $J_n$. We further Fourier transform the cavity amplitude to obtain the response in the frequency domain
\begin{equation}
    \alpha[\omega] = -2\pi \sqrt{\frac{\kappa_{\mathrm{ext}}}{2}} \; |\alpha_\mathrm{p}|
    \sum_{n,k} \frac{e^{i(n-k)\phi}  J_n(z_1) J_k(z_1) \delta(\omega+(n-k)\Omega_\mathrm{m})}{i(-\Delta+\mathcal{K}_{\mathrm{eff}}\bar{n}_{\mathrm{c}}+n\Omega_\mathrm{m})+\kappa/2} ,
\end{equation}
where $J_n$ denotes the Bessel function of the first kind. The solution in the frequency domain clearly shows that the cavity response has contributions from Bessel functions at multiples of mechanical resonance frequency $\Omega_{\mathrm{m}}$. As in Eq.~\ref{si:scatresponse1}, we set $\omega = 0$, since the transmission measurement is a single-tone experiment. Then we employ the input-output relation to obtain the scattering response of the microwave cavity in the unstable regime
\begin{equation}\label{si:cavityresponse}
    S_{21}[\omega=0] = 1- \frac{\kappa_{\text{ext}}}{2}
    \Sigma_{n} \frac{J_{n}(z_1) J_{n}(z_1)}{-i(\Delta+\mathcal{K}_{\text{eff}}n_{\text{c}}-n\Omega_{\text{m}})+\kappa/2} . 
\end{equation}
Since our ansatz in Eq.~\ref{si:eq:ansatz-mechannics-beta} introduces a parameter $B$, we fix the value of this oscillation amplitude by using the standard power balance equation \cite{marquardt2006,Ludwig2008,aspelmeyer-cavity-optomechanics}
\begin{equation}\label{si:powerbalanceeq}
   \Gamma_\mathrm{m}+\Gamma_{\text{opt}}(B)=0 ,
\end{equation} 
where the optomechanically induced damping rate $\Gamma_{\mathrm{opt}}(B)$ is calculated using the equation of motion for the oscillating part of the mechanical mode \cite{Armour2012}. Using Eq.~\ref{mechanical-mode-eq} and our ansatz for the mechanical mode as in Eq.~\ref{si:eq:ansatz-mechannics-beta}, we can write the following equation for the time evolution of the oscillation amplitude \cite{rodrigues-2010}
\begin{equation}
    \dot{B} = - \frac{(\Gamma_\mathrm{m}+\Gamma_{\mathrm{opt}})}{2} B,
\end{equation}
where $\Gamma_{\mathrm{opt}}$ is given by
\begin{equation}
    \Gamma_{\mathrm{opt}}(B) = \frac{2g_0 |\alpha|^2 \sin{(\omega_m t+\phi)}}{B}.
    \label{si:eq:gamma-opt-of-B}
\end{equation}
Here, $\Gamma_{\mathrm{opt}}$ depends on the optomechanical coupling $g_0$ and the cavity mode solution $\alpha$. Plugging Eq.~\ref{cavity-sol-time} into the expression for $\Gamma_{\mathrm{opt}}(B)$ in Eq.~\ref{si:eq:gamma-opt-of-B} and averaging over one mechanical time cycle gives
\begin{equation}
 \Gamma_{\mathrm{opt}}(B)=\frac{ g_0 x_{\mathrm{zpf}} \kappa_{\mathrm{ext}} n_{\mathrm{in}}}{2B} \mathbf{Im} \left[ \sum_n \frac{J_n(z_1) J_{n+1}(z_1)}{[i(\Delta+\mathcal{K}_{\mathrm{eff}}\bar{n}_{\mathrm{c}}-n\Omega_\mathrm{m})+\kappa/2][-i(\Delta+\mathcal{K}_{\mathrm{eff}}\bar{n}_{\mathrm{c}} - (n+1)\Omega_\mathrm{m})+\kappa/2]}\right]. 
\end{equation}
In addition, the photon number equation also gets modified in the unstable regime. There, it can be obtained by multiplying Eq.~\ref{cavity-sol-time} with its complex conjugate and averaging over one mechanical cycle \cite{ludwig2013}
\begin{equation}
\bar{n}_{\mathrm{c}} = \frac{\kappa_{\mathrm{ext}}}{2} n_{\mathrm{in}} \sum_n \frac{J_n(z_1) J_n(z_1)}{(\Delta+\mathcal{K}_{\mathrm{eff}}\bar{n}_{\mathrm{c}}-n\Omega_\mathrm{m})^2+\kappa^2/4} .  
\label{si:eq:n_c_unstable}
\end{equation}
As compared to Eq.~\ref{photon-no-eq}, which describes the photon occupation for the stable system, additional contributions appear from the Bessel functions and can be associated with the large amplitude oscillations of the mechanics. The photon number equation given in Eq.~\ref{si:eq:n_c_unstable} above, along with Eq.~\ref{si:powerbalanceeq}, forms a self-consistent set of equations, which is iteratively solved to obtain $\bar{n}_{\mathrm{c}}$ and $B$. The resulting values are then plugged into Eq.~\ref{si:cavityresponse} to obtain the scattering response in the unstable regime.

\subsection{Simulation}\label{si:simulation}

An alternative route to obtain the cavity response is to numerically solve the equations of motion and use the standard input-output relation. We start with the nonlinear equations of motion given in Eqs.~\ref{cavity-mode-eq} and \ref{mechanical-mode-eq}, and rewrite them in terms of the real and imaginary parts of the complex amplitudes $(\alpha_\mathrm{r}, \alpha_\mathrm{i}, \beta_\mathrm{r}, \beta_\mathrm{i})$ using the following relations
\begin{equation}
    \alpha_{\mathrm{r}} = \frac{\alpha+\alpha^*}{2}, \alpha_{\mathrm{i}} = \frac{i(\alpha^*-\alpha)}{2}, \beta_{\mathrm{r}} = \frac{\beta+\beta^*}{2}, \text{and} \; \beta_{\mathrm{i}} =\frac{i(\beta^*-\beta)}{2}.
\end{equation}
This allows us to express the equations in terms of real quantities since $\alpha_{\mathrm{r}} (\beta_{\mathrm{m}})$ and $\alpha_{\mathrm{i}} (\beta_{\mathrm{i}})$ are proportional to the position and momentum of the cavity (mechanics). Following are the equations of motion when expressed in the modified basis:
\begin{subequations}\label{si:eqsquadratures}
\begin{align}
    \mathrm{\frac{d}{dt}} \alpha_\mathrm{r} &= -\Delta \alpha_\mathrm{i} - \frac{\kappa}{2}\alpha_\mathrm{r} - \mathcal{K} (\alpha_\mathrm{r}^2+\alpha_\mathrm{i}^2)\alpha_\mathrm{i} + 2g_0 \beta_\mathrm{r}\alpha_\mathrm{i} - \sqrt{\frac{\kappa_{\mathrm{ext}}}{2}}\alpha_\mathrm{p},\\
    \mathrm{\frac{d}{dt}} \alpha_\mathrm{i} &= \Delta \alpha_\mathrm{r} - \frac{\kappa}{2} \alpha_\mathrm{i} + \mathcal{K} \alpha_\mathrm{r}(\alpha_\mathrm{r}^2+\alpha_\mathrm{i}^2) - 2g_0 \beta_\mathrm{r}\alpha_\mathrm{r},\\
    \mathrm{\frac{d}{dt}} \beta_\mathrm{r} &= \Omega_\mathrm{m} \beta_\mathrm{i} - \frac{\Gamma_\mathrm{m}}{2} \beta_\mathrm{r},\\
    \mathrm{\frac{d}{dt}} \beta_\mathrm{i} &= -\Omega_\mathrm{m} \beta_\mathrm{r} -\frac{\Gamma_\mathrm{m}}{2} \beta_\mathrm{i} - g_0 (\alpha_\mathrm{r}^2+\alpha_\mathrm{i}^2).
\end{align}
\end{subequations}
We choose to model the mean field equations without using explicit noise terms because the $S_{21}$ measurement performed in the experiment is single-tone, which mainly gives access to the mean response of the setup and can be modeled without studying the quantum fluctuations. We solve this coupled set of differential equations for different values of the detuning $\Delta$ and the input power $n_{\mathrm{in}}$. This allows us to precisely predict the timescale required for the system to reach its steady state. Furthermore, it provides us with another possible method to obtain the oscillation amplitude of the mechanical mode, since the oscillation amplitude is simply given by $B = |\beta_\mathrm{r}^2 + \beta_\mathrm{i}^2| x_{\mathrm{zpf}}$.

The code used to obtain the numerical solution is provided and can be used to reproduce the simulation results in Figure \ref{main:fig4:colormap-slices}. The numerical simulation does not include the approximations used for the analytical model, such as assuming the steady state to be able to calculate $\bar{n}_{\mathrm{c}}$, thus allowing us to account for the correct photon number. Besides, since our system operates in the bistable regime, the numerical solution automatically chooses the correct photon number branch, whereas in the analytical model, we need to specify the photon branch that we are in. Using the input-output relation in the time domain
\begin{equation}
    \alpha_{\mathrm{out}}(t) = \alpha_{\mathrm{in}}(t) + \sqrt{\frac{\kappa_{\mathrm{ext}}}{2}} \alpha(t),
\end{equation}
the time-dependent scattering response of the microwave cavity $S_{21}(t)$ is obtained as
\begin{equation}
    S_{21}(t) = 1 + \sqrt{\frac{\kappa_{\mathrm{ext}}}{2}} \frac{1}{\alpha_\mathrm{p}} (\alpha_\mathrm{r}(t) +i \alpha_\mathrm{i}(t)) .
\end{equation}
Here, $\alpha_{\mathrm{in}} = \alpha_\mathrm{p}$ represents the input drive power. Since the system is self-oscillating, both the mechanical and the cavity amplitude oscillates around a mean value, which is measured in the experiment. We ensure that we choose a time interval which is larger than the time period of the oscillations, and take an average to obtain the steady state value of $|S_{21}|$ for a given set of parameters ($\Delta$, $n_{\mathrm{in}}$). This is equivalent to taking the zero-frequency component in the frequency domain, i.e. setting $\omega=0$ in the analytical model. It can be understood in the following manner: For a continuous signal $f(t)$, the Fourier transform is defined as
\begin{equation}
    \mathcal{F}[\omega] = \int_{-\infty}^{\infty} f(t) e^{-i\omega t} \, \mathrm{d}t ,
\end{equation}
In our case, since we have a finite and discrete time domain signal, $ x[t] $ with $ t = 0, 1, \dots, N-1 $, we need to apply a Discrete Fourier Transform (DFT), which is defined as
\begin{equation}
X[\omega] = \frac{1}{N}\sum_{t=0}^{N-1} x[t] \, e^{-i 2\pi \omega t / N}, \quad k = 0, 1, \dots, N-1.
\end{equation}
For the DC or zero frequency component (when \( \omega = 0 \)), the expression simplifies to
\begin{equation}
X[0] = \frac{1}{N}\sum_{t=0}^{N-1} x[t] = \langle x \rangle,
\end{equation}
where $\langle x \rangle$ is the average of the time domain signal. From that we can conclude that the zero-frequency (DC) component is equal to the average or constant background level of the signal. Thus, by removing the oscillatory components (those with non-zero frequency), we obtain the time-averaged component of the scattering response.

\begin{figure}[t]
\centering
   \includegraphics[scale=0.65]{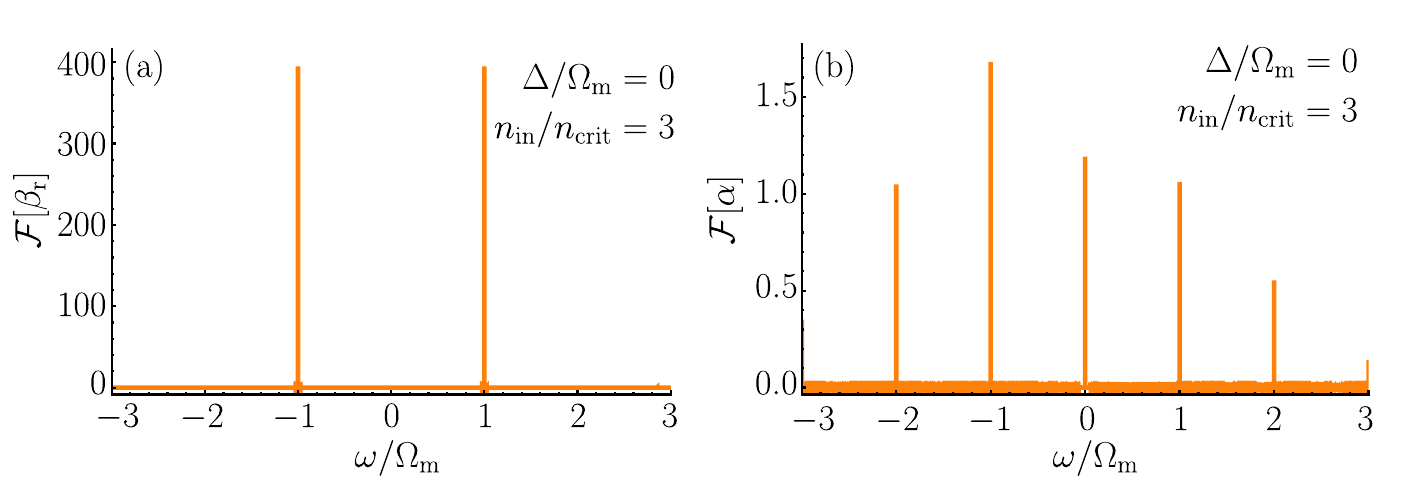}
   \caption{(a) The Fourier transform $\mathcal{F} \left[\beta_\mathrm{r}\right]$ of the mechanical mode contains frequency components only at $\pm \Omega_\mathrm{m}$. (b) The Fourier transform $\mathcal{F} \left[\alpha\right]$ of the cavity mode contains components at multiples of the mechanical frequency $\pm N \Omega_\mathrm{m}$ (where $N = 0, 1, 2, \dots$).}
    \label{app:fourier-transform}
\end{figure}

The simulation also offers a way to validate the assumptions made in the analytical description. There, we assumed an ansatz of the form $\beta = \bar{\beta}+\frac{B}{x_{\mathrm{zpf}}} e^{-i\Omega_\mathrm{m} t} e^{-i\phi}$ for the mechanical mode when it is oscillating at a high amplitude due to instabilities. This ansatz assumes that the mechanical oscillation is dominated by the contribution from the frequencies $\pm \Omega_\mathrm{m}$. We can verify this claim by calculating the Fourier transform of the numerical solution of $\beta_r$, which reveals the frequency components contributing to the oscillation. Fig.~\ref{app:fourier-transform} (a) shows the Fourier transform of the displacement of the mechanical mode $\beta_\mathrm{r}$. We observe that the only two frequencies contributing to the oscillation of the mechanical mode are $\pm \Omega_m$. In contrast, the cavity mode $\alpha$ shows additionally frequency components at multiples of the mechanical frequency, as shown in Fig.~\ref{app:fourier-transform} (b). This is a clear signature of self-sustained oscillations in the system \cite{marquardt2006}.

To summarize, we have obtained the scattering response of the cavity by both using analytical and numerical techniques in the stable and unstable regime. In the next section, we perform a linear stability analysis to determine the regions of the phase space where the system is unstable. We also determine the different kinds of bifurcations in the phase diagram.

\section{Stability analysis} \label{si:phasediagram}

The stability analysis is performed by first identifying the fixed points of the system, which are points at which $(\dot{\alpha}_\mathrm{r}, \dot{\alpha}_\mathrm{i}, \dot{\beta}_\mathrm{r}, \dot{\beta}_{\mathrm{i}}) = 0$. For our setup, this implies setting the right-hand side of the equations of motion given in Eqs.~\ref{si:eqsquadratures} to zero. We denote the solutions of the fixed points as $(\alpha_\mathrm{r}^*, \alpha_\mathrm{i}^*, \beta_\mathrm{r}^*, \beta_\mathrm{i}^*)$ for the cavity and mechanical mode, respectively. After determining the fixed points or equilibrium solutions, we compute the Jacobian matrix from the equations of motion (Eqs.~\ref{si:eqsquadratures}) as well to obtain the eigenvalues around each fixed point. These determine the stability of the given fixed point \cite{Strogatz-nonlin-dyn-chaos}. The Jacobian matrix for our setup is given by
\begin{equation}
     J = \begin{pmatrix}
        -\kappa/2 -2\mathcal{K} \alpha_\mathrm{r} \alpha_\mathrm{i} & -\Delta-3\mathcal{K}\alpha_\mathrm{i}^2 +2g_0\beta_\mathrm{r} -\mathcal{K}\alpha_\mathrm{r}^2&
        2g_0\alpha_\mathrm{i} &0 \\
        \Delta +3\mathcal{K}\alpha_\mathrm{r}^2+\mathcal{K}\alpha_\mathrm{i}^2-2g_0\beta_\mathrm{r}&
        -\kappa/2 + 2\mathcal{K} \alpha_\mathrm{r} \alpha_\mathrm{i} & -2g_0\alpha_\mathrm{r} & 0 \\
        0&
        0 &
        -\Gamma_\mathrm{m}/2 &\Omega_\mathrm{m}
         \\
        -2g_0\alpha_\mathrm{r} &
        -2g_0\alpha_\mathrm{i} &
        -\Omega_\mathrm{m} &
        -\Gamma_\mathrm{m}/2
    \end{pmatrix} .
\end{equation}
In our nonlinear system, there can be either one or three solutions for the fixed points depending on the detuning and input power. The stability of each fixed point is checked by evaluating the eigenvalues of the above Jacobian matrix at that point. If all the eigenvalues of the Jacobian matrix have a negative real part, the corresponding fixed point is stable. In all other cases, it is unstable. This is done using the code provided in the manuscript, which can be used to reproduce the stability diagrams obtained in Figure\,\ref{main:fig2:stability-dia}.

\begin{figure}
    \centering
    \includegraphics[width=0.5\linewidth]{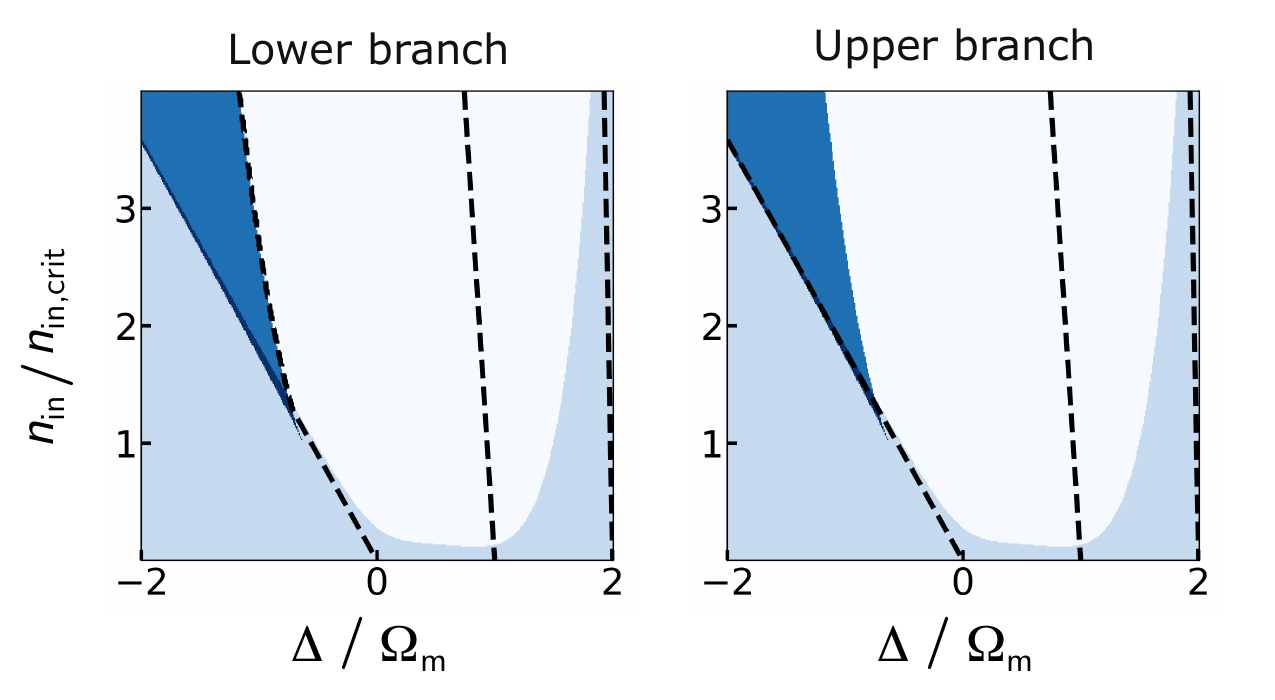}
    \caption{Stability diagram of set IV along with dashed lines corresponding to the effective shifted resonance ($\tilde{\Delta}=0$), the first blue sideband ($\tilde{\Delta}=\Omega_\mathrm{m}$) and the second blue sideband ($\tilde{\Delta}=2\Omega_\mathrm{m}$). The shift is governed by the low (left) and high (right) photon branch.}
    \label{si:fig:stability-dia}
\end{figure}

We have already looked at the stability diagram in Figure\,\ref{main:fig2:stability-dia} of the main text. However, we can understand the shifting of the unstable region in more detail by looking at the Kerr-shifted detuning $\tilde{\Delta}$, which defines the resonance and sideband condition in terms of the Kerr shifted resonance with respect to low and high photon branch. In Fig.~\ref{si:fig:stability-dia}, we plot the stability diagram for parameter set IV. We highlight the evolution of the resonance condition and the effective blue sidebands $\tilde{\Delta}=0,\Omega_\mathrm{m},2\Omega_\mathrm{m}$ with input power $n_\mathrm{in}$ as dashed lines. From Fig.~\ref{si:fig:stability-dia}, it is evident that only the resonance frequency, i.e.~$\tilde{\Delta}=0$, is strongly shifted by the Kerr nonlinearity with increasing input power $n_\mathrm{in}$. The effective first and second sidebands ($\tilde{\Delta}=\Omega_\mathrm{m},2\Omega_\mathrm{m}$) do not show a strong shift because the shift is proportional to the number of photons circulating in the cavity, which is small due to the low power operation regime and the detuning from the cavity resonance.

\begin{figure}[t]
    \centering
    \includegraphics[width=0.95\linewidth]{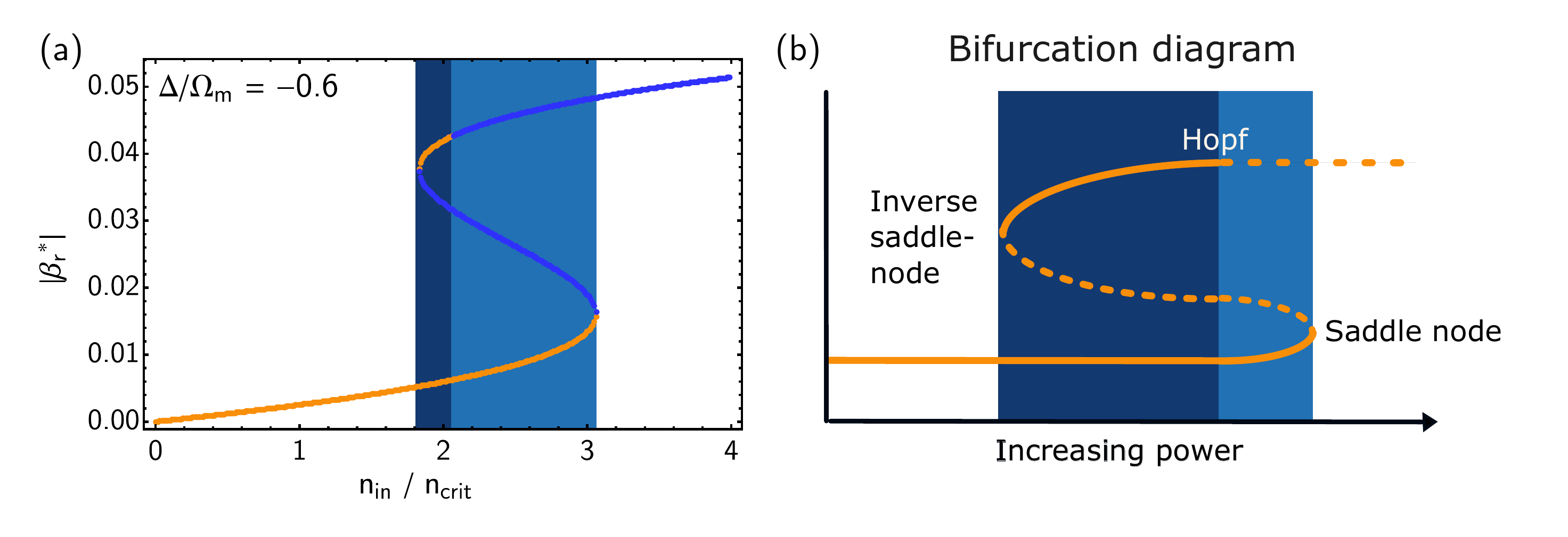}
    \caption{(a) Fixed points as a function of input power for parameter set III in table \ref{SI:tab:param-sets-all}. A blue color of the line denotes that the fixed point is unstable, whereas an orange color denotes that it is stable. (b) Qualitative bifurcation diagram. We show the evolution of the fixed point(s) with increasing probe power. Solid lines represent stable and dashed lines represent unstable fixed points.}
    \label{fixedpointvspower}
\end{figure}

Furthermore, by looking at the stable and unstable fixed points as a function of the input power, we qualitatively determine the types of bifurcations in the phase diagram. Generally, one can differentiate between two different types of bifurcations: 
\begin{enumerate}[label=(\roman*)]
    \item A saddle node (inverse saddle node) bifurcation, which is characterized by the annihilation (creation) of a pair of fixed points, consisting of one stable and one unstable fixed point.
    \item A Hopf bifurcation, which is marked by a change in the stability of a fixed point, where a stable fixed point becomes unstable as the bifurcation parameter (in this case $n_{\mathrm{in}}$) is changed, and a periodic orbit emerges (also known as limit cycle).
\end{enumerate}
To visualize the different types of bifurcation, we calculate the evolution of the fixed point $\beta_\mathrm{r}^*$ as the input power increases for a frequency where the system exhibits multi-stability. The result is shown Fig.~\ref{fixedpointvspower}(a), where the system transitions from a single fixed point to three fixed points, and finally goes back to a single fixed point again. Each of the points can be either stable (orange) or unstable (blue). In Fig.~\ref{fixedpointvspower}(b), we obtain a qualitative picture of the bifurcations in the multi-stable regime. Here, the system first undergoes an inverse saddle node bifurcation. This is followed by a Hopf bifurcation, where the stable fixed point becomes unstable and an isolated periodic orbit (limit cycle) arises in the phase space. Increasing the input power further leads to a saddle node bifurcation where a pair of a stable and unstable fixed point gets annihilated.

Since the fixed points predict the long-time temporal behavior of the system, they provide an alternate route to find the steady state photon occupation in the microwave cavity. The photon occupation given by Eq.~\ref{photon-no-eq} can be obtained by calculating $\alpha_\mathrm{r}^{*2}+\alpha_\mathrm{i}^{*2}$, where $\alpha_\mathrm{r}^* (\alpha_\mathrm{i}^*)$ denote the solutions to the fixed points for the real and imaginary part of the cavity mode. Similar to the cubic equation, this also gives us three possible solutions for the photon number, which correspond to the well-known three branches of the photon number obtained from the cubic equation (Eq.~\ref{photon-no-eq}).

\begin{figure}[t]
\centering
\includegraphics[scale=0.9]{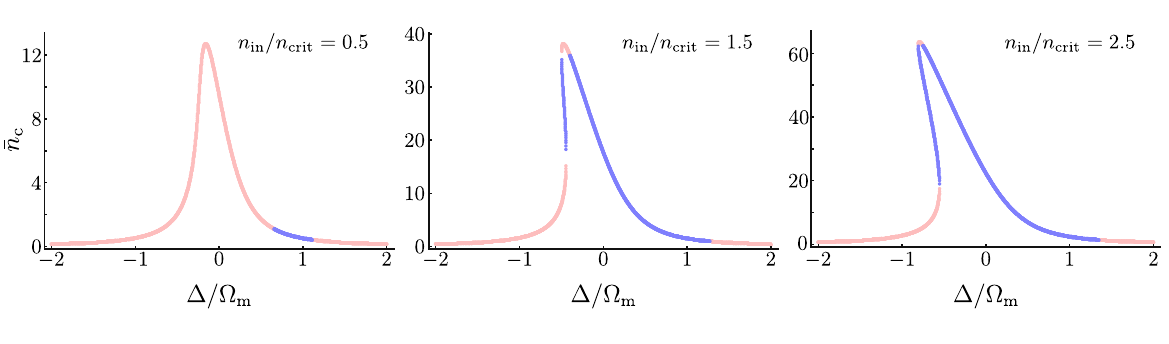}
    \caption{Photon number $\bar{n}_\mathrm{c}$ obtained from the fixed point solution (the light pink color corresponds to stable and the light purple to unstable solution(s)) as a function of detuning $\Delta$ normalized on the mechanical resonance frequency $\Omega_\mathrm{m}$ for different input powers from $n_\mathrm{in} / n_{\mathrm{in,crit}} = 0.5$ to $1.5$ in the middle and $2.5$ on the right. We exemplarily depict this for parameter set III in table \ref{SI:tab:param-sets-all} here. Above bifurcation ($n_\mathrm{in} / n_{\mathrm{in,crit}} \geq 1$), the solution shows three branches similar to the photon number branches obtained from the cubic equation (Eq.~\ref{photon-no-eq}). However, in addition to the middle branch, the upper branch is also unstable here due to the instability of the mechanical subsystem.}
    \label{app:photonbranch}
\end{figure}

Fig.~\ref{app:photonbranch} shows the photon occupation $\bar{n}_{\mathrm{c}}$ obtained from the fixed points as a function of the detuning for different input powers $n_\mathrm{in}$. The input power is chosen such that the system transitions to the bistable regime, as is expected from a nonlinear optomechanical system. The behavior of the photon number shows similarity to a Duffing nonlinearity, where the system exhibits three possible branches or solutions for the photon number, and the middle branch is unstable. However in our case, due to the backaction heating in the blue sideband, we see that the upper branch also becomes unstable and thus inaccessible in the experiment. This behavior is similar to a linear optomechanical system $(\mathcal{K}=0)$ driven into dynamical instability by high input driving powers \cite{aldana2013}.

\begin{figure}[t]
\centering
\includegraphics[scale=0.9]{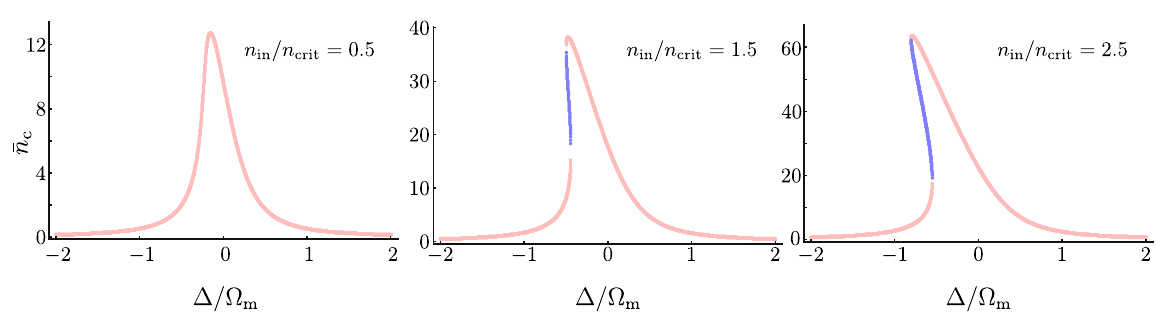}
    \caption{Photon number $\bar{n}_\mathrm{c}$ obtained from the fixed point solution (the light pink color corresponds to stable and the light purple to unstable solution(s)) for $g_0 / 1000$ reduced by three orders of magnitude when compared to Fig.~\ref{app:photonbranch}. All other parameters correspond to set III in table \ref{SI:tab:param-sets-all}. Due to the weaker optomechanical coupling, we recover the case of a Kerr nonlinear resonator with two stable branches (upper and lower) and one unstable branch (middle) above bifurcation.}
    \label{app:photonbranch-low-g0}
\end{figure}

The instability in the upper photon branch is attributed to the strong optomechanical coupling $g_0$. The backaction effects due to the strong $g_0$ can not only lead to heating of the mechanical mode, but they can also push the state beyond simple heating into the unstable regime, where the total damping rate of the mode becomes negative. Since the instability is created via the backaction in the optomechanical interaction, decreasing the optomechanical coupling rate $g_0$ decreases the backaction as well and thereby restores the stability of the upper photon branch. This can be seen in Fig.~\ref{app:photonbranch-low-g0}, which shows the photon occupation for $g_0 / 1000$. Above bifurcation, we clearly observe that, in this case, the upper and lower photon number branches of the system are both stable, while only the middle branch is unstable. This exactly matches the case of a classical Kerr nonlinear or a Duffing resonator, which is not optomechanically coupled to a mechanical oscillator.

In conclusion, our stability analysis not only allows to understand the regions of instability but also provides insight into other aspects, such as bifurcations.

\end{document}